\documentclass[paperpaper,twocolumn,english,superscriptaddress,aps,prx,floatfix,amsfonts,amssymb]{revtex4-1}

\usepackage{epsfig}
\usepackage{graphicx}
\usepackage{dcolumn}
\usepackage{bm}
\usepackage{yfonts}
\usepackage{eufrak}
\usepackage{amsmath}
\usepackage{amssymb}
\usepackage{tikz}
\usepackage{xcolor}
\usetikzlibrary{arrows.meta, positioning}

\newcommand{\EQ}{\begin{equation}}
\newcommand{\EN}{\end{equation}}
\newcommand{\ea}{\end{eqnarray}}
\newcommand{\ba}{\begin{eqnarray}}

\newcommand{\bear}{\begin{eqnarray}}
\newcommand{\ear}{\end{eqnarray}}

\begin{document}
\title{Exact results for the Hubbard model on bipartite lattices in spatial dimensions $d>1$: Seven theorems from the full 
[SU(2)$\times$SU(2)$\times$U(1)]/$\mathbb{Z}_2^2$ symmetry}


\author{J. M. P. Carmelo}
\affiliation{Center of Physics of University of Minho and University of Porto, LaPMET, P-4169-007 Oporto, Portugal}
\affiliation{CeFEMA, Instituto Superior T\'ecnico, Universidade de Lisboa, LaPMET, Av. Rovisco Pais, P-1049-001 Lisbon, Portugal}



\begin{abstract}
There are few exact results for the Hubbard model on bipartite lattices of spatial dimension $d>1$. Nevertheless, the Hubbard model with transfer 
integral $t$ and onsite repulsion $U$ on bipartite lattices with $N_a$ sites, such as the square, honeycomb, cubic, body-centered cubic, 
face-centered cubic, and diamond lattices, provides the simplest toy model for describing electronic correlations in many condensed-matter 
systems and is therefore a quantum problem of considerable physical interest. Most previous studies have assumed that, for finite onsite 
repulsion $U>0$, the global symmetry of the model is SO(4) = [SU(2)$\times$SU(2)]/$\mathbb{Z}_2$. However, the actual global symmetry 
is larger and given by [SU(2)$\times$SU(2)$\times$U(1)]/$\mathbb{Z}_2^2$, which may equivalently be written as 
[SO(4)$\times$U(1)]/$\mathbb{Z}_2$ or as SO(3)$\times$SO(3)$\times$U(1). The main goal of this paper is to extract physical insight, 
including further exact results for the Hubbard model on bipartite lattices with spatial dimension $d>1$, from its full global symmetry. 
The hidden U(1) symmetry beyond SO(4), which has been ignored in most previous studies, becomes an explicit symmetry within an 
exact quasiparticle representation of the model for $u=U/t>0$. The generator of this symmetry has eigenvalues 
$S_{\tau} \in \{0,{1\over 2}, 1, {3\over 2}, 2,\ldots,{N_a\over 2}\}$. Importantly, this exact quasiparticle representation naturally leads to 
the emergence of $N_s = (N_a - 2S_{\tau})$ physical spins $1/2$ and $N_{\eta} = 2S_{\tau}$ physical $\eta$-spins $1/2$, associated 
with the two SU(2) symmetries in [SU(2)$\times$SU(2)$\times$U(1)]/$\mathbb{Z}_2^2$. The spin and $\eta$-spin multiplet and singlet 
configurations of the $4^{N_a}$ exact energy and momentum eigenstates are generated from the vacuum by quasiparticle creation and 
annihilation operators. Seven exact theorems that provide new physical insight into the model are established. Overall, the exact 
framework based on physical spins and physical $\eta$-spins for the Hubbard model on bipartite lattices of spatial dimension $d>1$ 
offers a robust foundation for future studies of the model, as well as of the condensed-matter materials, such as cuprate superconductors, 
graphene and graphene-derived systems, and other quantum systems, that it describes.
\end{abstract}
\maketitle

\section{Introduction}
\label{SECI}

Over the years, the Hubbard model \cite{Gutzwiller_63,Hubbard_63,Hubbard_64}, characterized by a transfer 
integral $t$ and an on-site repulsion $U$, has gained increasing importance, as it plays a central role in a wide 
range of problems in condensed-matter physics and related areas of quantum many-body theory. 

In particular, the Hubbard model on bipartite lattices, such as the square, honeycomb, cubic, body-centered cubic (BCC), 
face-centered cubic (FCC), and diamond lattices, provides the simplest toy model for describing electronic correlations 
in many condensed-matter materials and ultracold atomic systems, and therefore constitutes a quantum problem 
of considerable physical interest.

For instance, and of particular importance, the Hubbard model on the square lattice has been widely used to 
describe the properties of the cuprate superconductors 
\cite{Anderson_87,Zhang_88,Lee_92,Dagotto_94,Hayden_96,Imada_98,Ho_21,Sorella_02,Damascelli_03,Maier_05,
Lee_06,Maier_08,Comanac_08,Markiewicz_10,Das_10,Markiewicz_10A,Gull_10,Scalapino_12,Carmelo_12,Seibold_12,Das_14,LeBlanc_15,Piazza_15,
Markiewicz_15,Schafer_21,Kumar_21,Martinelli_22,Singh_24,Wang_24,Singh_25,Bao_25,Schumm_25}. 
Similarly, the Hubbard model on the honeycomb lattice has been employed to describe electronic interactions in graphene 
and graphene-derived systems \cite{Sorella_92,Gonzalez_01,Neto_09,Kotov_12,Katsnelson_12,Assaad_13,Cao_18,Martinez_25}. 
Moreover, the Hubbard model on bipartite lattices has been experimentally realized using ultracold atomic systems \cite{Kohl_05,Jordens_08,Schneider_08,Strohmaier_10,Sensarma_10}.

For the Hubbard model on bipartite lattices other than the one-dimensional (1D) lattice, only a limited number of exact 
results are known; some of these are summarized below. The particle-hole transformation was originally introduced 
in Ref. \onlinecite{Hubbard_63}. Nagaoka's theorem \cite{Nagaoka_66} states that if the lattice is sufficiently connected, 
which is the case for bipartite lattices of spatial dimension $d>1$, then the ground state of the Hubbard model at 
$u=U/t=\infty$ with exactly one hole is fully spin polarized, with total spin $S_s = {1\over 2}(N_a - 1)$, and is 
unique up to the trivial $(2S_s+1)$-fold SU(2) degeneracy. It is therefore a ferromagnetic ground state, despite 
the interaction being purely repulsive. 

In Ref. \onlinecite{Lieb_89}, the uniqueness of the ground state at half-filling 
was rigorously established for the Hubbard model on bipartite lattices. In that work, the ground-state spin was 
shown to be given by $S_s = {1\over 2}(N_B - N_A)$, where $N_A$ and $N_B$, with $N_A \leq N_B$, denote 
the numbers of sites in sublattices $A$ and $B$, respectively. In that reference, as part of the proof technique 
(using particle-hole symmetry and the Perron-Frobenius theorem for positive and non-negative matrices 
\cite{Perron_07,Frobenius_12}), it was established that for the ground state at half-filling,
\begin{equation}
\langle \vec{\hat{S}}_{s,j} \cdot \vec{\hat{S}}_{s,j'} \rangle =
\begin{cases}
\ge 0, & \vec{r}_j \in A,\ \vec{r}_{j'} \in B, \\[4pt]
\le 0, & \vec{r}_j,\vec{r}_{j'} \in A \ \text{or}\ \vec{r}_j,\vec{r}_{j'} \in B \, ,
\end{cases}
\label{SS}
\end{equation}
where $\vec{\hat{S}}_{s,j}$ is the spin operator at site $j$,
which is the Hubbard-model analog of Marshall-Lieb-Mattis antiferromagnetic sign structure.
That reference also provided a rigorous mathematical 
foundation for the SU(2)$\times$SU(2) global symmetry associated with spin and $\eta$-spin.

The corresponding $\eta$-spin multiplets were derived and connected to representation theory in Ref. \onlinecite{Yang_89}, 
where exact energy and momentum eigenstates forming highest-weight states of $\eta$-spin were identified. In Ref. \onlinecite{Yang_90}, 
it was shown that particle-hole symmetry, together with the global spin SU(2) symmetry and the global $\eta$-spin SU(2) 
symmetry, gives rise to a global SO(4) = [SU(2)$\times$SU(2)]/$\mathbb{Z}_2$ symmetry for the Hubbard model on 
bipartite lattices. This SO(4) symmetry has been assumed to be the global symmetry of the model in most subsequent studies \cite{Demler_04,Masumizu_05,Hart_15,Mazurenko_17,Brown_17,Fava_20,Moca_23}.

The Shen-Qiu-Tian theorem \cite{Shen_94} refers to a set of rigorous results on the ground-state spin and $\eta$-spin
structure of the Hubbard model on bipartite lattices, especially concerning ferrimagnetic order and the connection 
between spin and lattice imbalance. These results generalize Lieb's earlier theorems \cite{Lieb_89} about ground-state 
quantum numbers of the Hubbard model.

The subgroups that constitute local gauge symmetries of the Hubbard model were identified in Ref. \onlinecite{Ostlund_91}. 
In that work, a ``non-local'' gauge U(1) symmetry of the model in the $u = U/t \rightarrow \infty$ limit was discovered, with 
the corresponding local gauge symmetry in this limit given by SU(2)$\times$SU(2)$\times$U(1). This local symmetry was 
subsequently promoted in Ref. \onlinecite{Carmelo_10} to a global [SU(2)$\times$SU(2)$\times$U(1)]/$\mathbb{Z}_2^2$ 
symmetry of the Hubbard model on any bipartite lattice for finite values of $u = U/t$.

That symmetry may be expressed equivalently as SO(3)$\times$SO(3)$\times$U(1) or as [SO(4)$\times$U(1)]/$\mathbb{Z}_2$, 
where the U(1) symmetry beyond SO(4) is associated with the relative translational degrees of freedom of the spin SU(2) and 
$\eta$-spin SU(2) symmetries, respectively. It is therefore not solely related to the charge degrees of freedom, in contrast to the 
preliminary interpretation presented in Ref. \onlinecite{Carmelo_10}. In this paper, we refer to this symmetry as the 
$\tau$-translational U(1) symmetry, whose generator has eigenvalues $S_{\tau} \in \{0,{1\over 2}, 1, {3\over 2}, 2,\ldots,{N_a\over 2}\}$.

The studies presented in this paper focus on the Hubbard model on bipartite lattices other than the 1D lattice, 
for which an exact Bethe-ansatz solution exists \cite{Lieb_68,Lieb_03,Takahashi_72,Martins_97,Martins_98,Essler_05}. 
The quantum numbers of that solution have been related to the model's [SU(2)$\times$SU(2)$\times$U(1)]/$\mathbb{Z}_2^2$ 
global symmetry \cite{Carmelo_25}. The $\tau$-translational U(1) symmetry beyond SO(4) was shown in 
Refs. \onlinecite{Carmelo_24,Carmelo_25A,Carmelo_26} to play a key role in rendering charge transport diffusive in the 1D Hubbard 
model at half-filling and finite temperatures, rather than superdiffusive, as predicted by studies that accounted only for 
the two SU(2) symmetries in [SU(2)$\times$SU(2)$\times$U(1)]/$\mathbb{Z}_2^2$ \cite{Fava_20,Moca_23}.

Moreover, the 1D Hubbard model has been extensively used to describe 
the physics of quasi-one-dimensional materials \cite{Jerome_82,Baeriswyl_87,Seo_97,Bourbonnais_98,Carmelo_06A}, in 
which several non-perturbative properties of the exotic quantum liquid it describes \cite{Ogata_90,Frahm_90,Frahm_91,Carmelo_93,Carmelo_94,Carmelo_94A,Carmelo_94B,Voit_95,Gebhard_97,Carmelo_03,Carmelo_04} 
have been experimentally observed. These include dynamical properties \cite{Penc_96,Carmelo_05,Carmelo_06,Carmelo_18,Gebhard_00} 
as well as transport properties \cite{Carmelo_25,Shastry_90,Peres_97}, whose study has employed methods similar to those 
used for other integrable models \cite{Neto_94,Peres_99,Carmelo_15}.

On the other hand, as mentioned above, quantum problems of considerable physical interest for many condensed-matter materials include 
the Hubbard model on bipartite lattices of spatial dimension $d>1$, such as the square, honeycomb, cubic, BCC, FCC, and diamond lattices. 
Our study focuses on this lattice quantum model, with the main goal of extracting physical insight from the 
[SU(2)$\times$SU(2)$\times$U(1)]/$\mathbb{Z}_2^2$ global symmetry discovered in Ref. \onlinecite{Carmelo_10}.

Seven exact theorems that provide new physical insight into the model are established. Our results on the exact quasiparticle representation, 
together with the associated physical-spin and physical-$\eta$-spin framework that emerges for $u = U/t > 0$ from this representation 
and symmetry, contribute to a deeper understanding of the physics of the Hubbard model on bipartite lattices with spatial dimensions $d > 1$. 
We anticipate that this physical-spin and physical-$\eta$-spin framework will be valuable for future studies, particularly those 
involving numerical simulations.

The paper is organized as follows. In Sec. \ref{SECII}, the global symmetry of the Hubbard model on bipartite lattices for $u = U/t > 0$ 
is revisited. Section \ref{SECIII} addresses the exact quasiparticle representation for spatial dimensions $d>1$. The emergence of 
$N_s = (N_a - 2S_{\tau})$ physical spins $1/2$ and $N_{\eta} = 2S_{\tau}$ physical $\eta$-spins $1/2$ from the quasiparticle 
representation is discussed in Sec. \ref{SECIV}. In Sec. \ref{SECV}, four exact theorems concerning the absence of level crossings in fully 
symmetry-resolved Hubbard Hamiltonian sectors, the absence of quasiparticle double occupancy 
and/or empty lattice sites in ground states, the extremal-weight 
structure of the ground state in external fields, and the correspondence between the number of ground states and irreducible 
representations are rigorously proved. The relation between the linear extent, the number of lattice sites, and the allowed momenta 
is also discussed in that section. An exact theorem on the exact factorization of internal SU(2) degrees of freedom and the localization 
of spin-charge coupling is rigorously established in Sec. \ref{SECVI}. In Sec. \ref{SECVII}, the differences with respect to the 1D Hubbard 
model are discussed. Two exact theorems on the proportionality of spin and $\eta$-spin currents in fixed $S_{\tau}$ sectors, as well 
as the vanishing of spin and $\eta$-spin currents in singlet sectors at fixed $S_{\tau}$, are rigorously proved in Sec. \ref{SECVIII}. 
Finally, Sec. \ref{SECIX} presents the concluding remarks and discussion. Three Appendixes contain additional results that are useful 
for the studies presented in this paper.

\section{The global symmetry for $U >0$}
\label{SECII}

For finite interaction $U>0$, the Hubbard model describes electronic correlations arising from the on-site repulsion 
among $N = (N_{\uparrow} + N_{\downarrow})$ electrons with spin projection $\sigma = \uparrow,\downarrow$. We 
employ natural units in which the Planck constant, the electronic charge, and the lattice spacing are set equal to one, 
and we assume the number of lattice sites $N_a$ to be an even integer. We consider $N_a$ to be very large, with the 
thermodynamic limit $N_a \rightarrow \infty$ being the regime of primary interest for real condensed-matter materials.

The Hamiltonian of the Hubbard model on a bipartite lattice $\Lambda\subset\mathbb{Z}^d$ 
with periodic boundary conditions is given by,
\begin{equation}
\hat{H}  = \hat{T} + U\hat{W} \, ,
\label{H}
\end{equation}
where
\begin{eqnarray}
\hat{T}  & = & -t\sum_{\substack{\langle j,j'\rangle}}
\sum_{\sigma}\Bigl(c_{\vec{r}_{j},\sigma}^\dagger c_{\vec{r}_{j'},\sigma} + {\rm h.c.}\Bigr)
= \sum_{l=0,\pm 1}\hat{T}_l \, \,
\nonumber \\
\hat{W} & = & \sum_{j=1}^{N_a}\hat{\rho}_{\vec{r}_j,\uparrow}\hat{\rho}_{\vec{r}_j,\downarrow}
= {1\over 2}\left(\hat{Q}_{D,E} - {1\over 2}\right) \, ,
\label{TU}
\end{eqnarray}
$c_{\vec{r}_j,\sigma}^{\dagger}$ creates one electron of spin projection $\sigma$ at site $j$,
\begin{eqnarray}
\hat{T}_{-1} & = & -t \sum_{\substack{\langle j,j'\rangle}}
\sum_{\sigma}(1 - \hat{n}_{\vec{r}_j,-\sigma})\,c_{\vec{r}_j,\sigma}^{\dagger}\,c_{\vec{r}_{j'},\sigma}\,\hat{n}_{\vec{r}_{j'},-\sigma}
\nonumber \\
\hat{T}_{+1} & = & -t \sum_{\substack{\langle j,j'\rangle}}
\sum_{\sigma}\hat{n}_{\vec{r}_j,-\sigma}\,c_{\vec{r}_j,\sigma}^{\dagger}\,c_{\vec{r}_{j'},\sigma}\,(1 - \hat{n}_{\vec{r}_{j'},-\sigma}) 
\nonumber \\
\hat{T}_0 & = & -t \sum_{\substack{\langle j,j'\rangle}}
\sum_{\sigma}\Bigl[\hat{n}_{\vec{r}_j,-\sigma}\,c_{\vec{r}_j,\sigma}^{\dagger}\,c_{\vec{r}_{j'},\sigma}\,\hat{n}_{\vec{r}_{j'},-\sigma}
\nonumber \\
& + & (1 - \hat{n}_{\vec{r}_j,-\sigma})\,c_{\vec{r}_j,\sigma}^{\dagger}\,c_{\vec{r}_{j'},\sigma}\,(1 - \hat{n}_{\vec{r}_{j'},-\sigma})\Bigr] \, ,
\label{T011}
\end{eqnarray}
are kinetic operators that change lattice-site electron double occupancy by $-1$, $1$, and $0$, respectively, 
$\hat{\rho}_{\vec{r}_j,\sigma} = (\hat{n}_{\vec{r}_j,\sigma} - 1/2)$, $\hat{n}_{\vec{r}_j,\sigma} = 
c_{\vec{r}_j,\sigma}^{\dagger}\,c_{\vec{r}_j,\sigma}$, and the operator,
\begin{equation}
\hat{Q}_{D,E} = \sum_{j=1}^{N_a}\Bigl[1 - \sum_{\sigma}\hat{n}_{\vec{r}_j,\sigma}(1 - \hat{n}_{\vec{r}_j,-\sigma})\Bigr] \, ,
\label{QED}
\end{equation}
counts the number of sites doubly occupied by electrons plus sites unoccupied by electrons.

The momentum operator is given by,
\begin{eqnarray}
\vec{\hat{P}} & = & \sum_{\vec{k},\sigma} c_{\vec{k},\sigma}^{\dagger}c_{\vec{k},\sigma}
\nonumber \\
& = & {1\over L}\sum_{j,j',\sigma}\Bigl(\sum_{\vec{k}} e^{i \vec{k}\cdot (\vec{r}_j - \vec{r}_{j'})}\Bigr)c_{\vec{r}_j,\sigma}^{\dagger}c_{\vec{r}_{j'},\sigma} \, .
\label{P}
\end{eqnarray}

At $U=0$, the global symmetry of the Hubbard model on a bipartite lattice, Eq. (\ref{H}), is
O(4)/$\mathbb{Z}_2$ = SO(4)$\times\mathbb{Z}_2$ \cite{Carmelo_10}. Here, the factor $\mathbb{Z}_2$ corresponds 
to the particle-hole transformation on a single spin, under which the interacting term of the Hamiltonian, Eq. (\ref{H}), 
is not invariant \cite{Ostlund_91}. The requirement that the global symmetry generators commute with the interacting 
Hamiltonian at $u = U/t \neq 0$ modifies this $U=0$ global symmetry to 
[SU(2)$\times$SU(2)$\times$U(1)]/$\mathbb{Z}_2^2$ = [SO(4)$\times$U(1)]/$\mathbb{Z}_2$, 
rather than reducing it to SO(4) alone \cite{Carmelo_10}. 

The well known generators of the global spin SU(2) symmetry and global $\eta$-spin SU(2) symmetry are given by,
\begin{eqnarray}
\hat{S}_s^{+} & = & \sum_{j=1}^{N_a}c_{\vec{r}_j,\downarrow}^{\dagger}c_{\vec{r}_j,\uparrow} \, ,
\hspace{0.20cm}
\hat{S}_s^{-} = \sum_{j=1}^{N_a}c_{\vec{r}_j,\uparrow}^{\dagger}c_{\vec{r}_j,\downarrow} \, ,
\nonumber \\
\hat{S}_s^z & = & {1\over 2}\sum_{j=1}^{N_a}(\hat{n}_{\vec{r}_j,\uparrow} - \hat{n}_{\vec{r}_j,\downarrow}) \, ,
\label{GeneS}
\end{eqnarray}
and
\begin{eqnarray}
\hat{S}_{\eta}^{+} & = & - \sum_{j=1}^{N_a}e^{i \vec{k}_{\eta}\vec{r}_j}c_{\vec{r}_j,\downarrow}^{\dagger}c_{\vec{r}_j,\uparrow}^{\dagger} \, ,
\hspace{0.20cm}
\hat{S}_{\eta}^{-} = - \sum_{j=1}^{N_a}e^{-i \vec{k}_{\eta}\vec{r}_j}c_{\vec{r}_j,\uparrow}c_{\vec{r}_j,\downarrow}
\nonumber \\
\hat{S}_{\eta}^z & = & {1\over 2}\sum_{j=1}^{N_a}(1 - \hat{n}_{\vec{r}_j,\uparrow} - \hat{n}_{\vec{r}_j,\downarrow}) \, ,
\label{GeneSeta}
\end{eqnarray}
respectively. Our choice of signs for the diagonal generators implies that the spin highest-weight states and the 
$\eta$-spin highest-weight states are transformed by the off-diagonal generators 
$\hat{S}_s^{+}$ and $\hat{S}_{\eta}^{+}$, respectively.

In the case of the off-diagonal generators 
$\hat{S}_{\eta}^{+}$ and $\hat{S}_{\eta}^{-}$, the wave vector $\vec{k}_{\eta}$ 
appearing in Eq. (\ref{GeneSeta}) is for any bipartite lattice, split into sublattices A and B, such that,
\begin{eqnarray}
e^{i \vec{k}_{\eta}\cdot\vec{r}_j} & = & +1\hspace{0.20cm}{\rm for}\hspace{0.20cm}\vec{r}_j \in A
\nonumber \\
& = & -1\hspace{0.20cm}{\rm for}\hspace{0.20cm}\vec{r}_j \in B \, ,
\label{keta}
\end{eqnarray}
with the constraint,
\begin{equation}
e^{2i\vec{k}_{\eta}\cdot\vec{r}_j} = 1\hspace{0.20cm}\Rightarrow\hspace{0.20cm}e^{i\vec{k}_{\eta}\vec{r}_j} = \pm 1 \, .
\label{keta2}
\end{equation}

Examples for several bipartite lattices of physical interest of coordinates of the vector $\vec{k}_{\eta}$ are,
\begin{eqnarray}
\vec{k}_{\eta} & = & (\pi,\pi) \rightarrow {\rm square}\hspace{0.20cm}{\rm lattice}
\nonumber \\
& = & \left({4\pi\over 3},0\right)\hspace{0.20cm}{\rm or}\hspace{0.20cm}\left({2\pi\over 3},{2\pi\over \sqrt{3}}\right)
\rightarrow {\rm honeycomb}\hspace{0.20cm}{\rm lattice}
\nonumber \\
& = & (\pi,\pi,\pi) \rightarrow {\rm cubic}\hspace{0.20cm}{\rm lattice}
\nonumber \\
& = & (2\pi,0,0) \rightarrow {\rm BCC}\hspace{0.20cm}{\rm lattice}
\nonumber \\
& = & (2\pi,2\pi,0) \rightarrow {\rm FCC}\hspace{0.20cm}{\rm lattice} 
\nonumber \\
& = & (2\pi,2\pi,2\pi) \rightarrow {\rm diamond}\hspace{0.20cm}{\rm lattice} \, ,
\label{severalketa}
\end{eqnarray}
where for the honeycomb lattice we have used wave vectors from the Brillouin zone corners,
for the BCC lattice the sites chosen are $A = (0,0,0)$ and $B = \left({1\over 2},{1\over 2},{1\over 2}\right)$,
for the FCC lattice $A = (0,0,0)$ and $B = \left({1\over 2},{1\over 2},0\right)$, and for the diamond lattice 
$A = (0,0,0)$ and $B = \left({1\over 4},{1\over 4},{1\over 4}\right)$.

The difficulty in identifying the $\tau$-translational U(1) symmetry beyond SO(4) in [SO(4)$\times$U(1)]/$\mathbb{Z}_2$ 
arises from the fact that it is a hidden symmetry. To clarify this issue, we construct, from any of the $4^{N_a}$ exact 
finite-$u$ energy and momentum eigenstates $\vert\Psi,u\rangle$ of the Hubbard model on a bipartite lattice, a corresponding energy 
and momentum eigenstate $\vert\Psi,\infty\rangle$ in the limit $u \rightarrow \infty$ as,
\begin{equation}
\vert\Psi,\infty\rangle = {\hat{V}}_u\vert\Psi,u\rangle \hspace{0.20cm}{\rm and}\hspace{0.20cm}
\vert\Psi,u\rangle = {\hat{V}}_u^{\dagger}\vert\Psi,\infty\rangle \, .
\label{Phistates}
\end{equation} 
Here, $\vert\Psi,u\rangle$ and $\vert\Psi,\infty\rangle$ have exactly the same $u$-independent quantum numbers 
and correspond to the same energy eigenstate, yet quantities such as the energy eigenvalues, electron double-occupancy 
average values, and the wave function differ for different values of $u = U/t$.

The unitary operator $\hat{V}_u$ can be written as,
\begin{equation}
{\hat{V}}_u^{\dagger} = e^{\hat{S}_u} \, , \hspace{0.20cm} {\hat{V}}_u = e^{-\hat{S}_u} 
\hspace{0.20cm}{\rm where}\hspace{0.20cm}
\lim_{u \to u_0}\| \hat V_u - \hat V_{u_0} \| = 0 
\label{hatVS}
\end{equation}
for $u>0$, so that it forms a strongly continuous unitary family.
The branch of $\hat{S}_u = \ln {\hat{V}}_u^{\dagger}$ is that for which the operator $\hat{S}_u$
is anti-Hermitian, $\hat{S}_u^{\dagger} = - \hat{S}_u$. 
Although the exact form of the anti-Hermitian operator $\hat{S}_u$ is not known and depends 
on the expression in terms of quasiparticle operators of the energy and momentum eigenstate 
$\vert\Psi,u\rangle$'s generators on the vacuum, 
as well as on the geometry of the bipartite lattice, an important exact result that follows from 
symmetry alone is that it involves {\it only} the kinetic operators $\hat{T}_{-1}$, $\hat{T}_{+1}$, 
and $\hat{T}_{0}$ in Eq. (\ref{T011}), with coefficients that depend on $u = U/t$ \cite{Carmelo_10}.

Its large-$u$ dominant term has a universal form given by \cite{Carmelo_10},
\begin{equation}
\hat{S}_u = - {1\over U}[\hat{T}_{+1} - \hat{T}_{-1}] + {\cal{O}} (t^2/U^2) \hspace{0.20cm}{\rm for}\hspace{0.20cm} u = U/t \gg 1 \, .
\label{Sularge}
\end{equation}
Therefore, the $u$-unitary operator ${\hat{V}}_u$ becomes the unit operator in the $u\rightarrow\infty$ limit.

Due to the highly degenerate spin and $\eta$-spin configurations in the $u \rightarrow \infty$ limit, there are many 
possible choices for a set of $4^{N_a}$ $u \rightarrow \infty$ energy and momentum eigenstates of the Hubbard 
model on a bipartite lattice \cite{Carmelo_10}.

A uniquely defined choice corresponds to the $4^{N_a}$ $u \rightarrow \infty$ energy and momentum eigenstates generated 
from the $4^{N_a}$ finite-$u$ energy and momentum eigenstates, as given in Eq. (\ref{Phistates}). Only this basis of $4^{N_a}$ 
exact $u \rightarrow \infty$ energy and momentum eigenstates is relevant for our study.

As with all other choices of $u \rightarrow \infty$ energy and momentum eigenstates, the lattice electronic occupancy 
configurations of the specific states $\vert\Psi,\infty\rangle$, Eq. (\ref{Phistates}), are such that the numbers 
of sites singly occupied by electrons with spin projection $\sigma = \uparrow$ and $\sigma = \downarrow$, 
sites unoccupied by electrons, and sites doubly occupied by electrons are good quantum numbers.

Although these numbers are not good quantum numbers at finite $u = U/t$, the construction of the exact 
many-electron energy and momentum eigenstates $\vert\Psi,u\rangle$ at each finite value of $u = U/t$ involves the transformation 
performed by the $u$-unitary operator ${\hat{V}}_u$ 
associated with quasiparticles for which they constitute good quantum numbers.

The $u = U/t > 0$ global $\tau$-translational U(1) symmetry beyond SO(4) in [SO(4)$\times$U(1)]/$\mathbb{Z}_2$ 
is a hidden symmetry because the expression of its generator,
\begin{eqnarray}
&& \hat{S}_{\tau} = {1\over 2} {\hat{V}}_u^{\dagger}\hat{Q}_{D,E}{\hat{V}}_u =
\nonumber \\
&& {1\over 2}(\hat{Q}_{D,E} + [\hat{Q}_{D,E},\hat{S}_u] + {1\over 2}[ [\hat{Q}_{D,E},\hat{S}_u],\hat{S}_u] + ... ) \, ,
\label{Gentau}
\end{eqnarray}
contains an infinite number of terms when expressed in terms of electron creation and annihilation operators. 

The operator $\hat{Q}_{D,E}$ appearing in Eq. (\ref{Gentau}) is defined in Eq. (\ref{QED}), and we have used the 
Baker-Campbell-Hausdorff expansion to express $\hat{S}_{\tau}$ in terms of electron creation and annihilation operators.
(In Ref. \onlinecite{Carmelo_10}, the operator counting the number of sites singly occupied by electrons replaces $\hat{Q}_{D,E}$ 
in the generator expression, Eq. (\ref{Gentau}). However, both choices of the generator of the global $\tau$-translational 
$U(1)$ symmetry are valid.)

Although the algebra involved in deriving the commutators of the three kinetic operators $\hat{T}_{-1}$, $\hat{T}_{+1}$, 
and $\hat{T}_{0}$ in Eq. (\ref{T011}) with the momentum operator, Eq. (\ref{P}), and the
six generators in Eqs. (\ref{GeneS}) and (\ref{GeneSeta}) is cumbersome, it is straightforward. 

All these commutators are found to vanish. Since the expression of the anti-Hermitian operator 
$\hat{S}_u$ in ${\hat{V}}_u = e^{-\hat{S}_u}$ involves only the kinetic operators $\hat{T}_{-1}$, $\hat{T}_{+1}$, and 
$\hat{T}_{0}$ in Eq. (\ref{T011}) with coefficients that depend on $u = U/t$, the following commutators also vanish exactly,
\begin{eqnarray}
&& [\vec{\hat{P}},{\hat{V}}_u] = [\hat{S}_s^{+},{\hat{V}}_u] = [\hat{S}_s^{-},{\hat{V}}_u] = [\hat{S}_s^z,{\hat{V}}_u] = 0 
\nonumber \\
&& [\hat{S}_{\eta}^{+},{\hat{V}}_u] = [\hat{S}_{\eta}^{-},{\hat{V}}_u] = [\hat{S}_{\eta}^z,{\hat{V}}_u] = 0 \, .
\label{SSScomm}
\end{eqnarray}

The hidden character of the global $\tau$-translational $U(1)$ symmetry beyond $SO(4)$ is consistent
with its generator $\hat{S}_{\tau}$, Eq. (\ref{Gentau}), not commuting with ${\hat{V}}_u$ except in the $u = U/t \rightarrow \infty$ limit,
when it becomes a gauge U(1) symmetry,
\begin{eqnarray}
[\hat{S}_{\tau},{\hat{V}}_u] & \neq & 0 \hspace{0.20cm}{\rm for}\hspace{0.20cm}{\rm finite}\hspace{0.20cm}u = U/t
\nonumber \\
& = & 0 \hspace{0.20cm}{\rm for}\hspace{0.20cm}u = U/t \rightarrow \infty \, .
\label{Staucomm}
\end{eqnarray}

The global [SU(2)$\times$SU(2)$\times$U(1)]/$\mathbb{Z}_2^2$ symmetry of the Hubbard model on a bipartite lattice is associated 
with the vanishing of the following commutators involving its Hamiltonian, Eq. (\ref{H}) \cite{Carmelo_10},
\begin{equation}
[\hat{H},\hat{S}_{\tau}] = [\hat{H},\vec{\hat{S}}_s^2] = [\hat{H},\hat{S}_s^z] =
[\hat{H},\vec{\hat{S}}_{\eta}^2] = [\hat{H},\hat{S}_{\eta}^z] = 0 \, .
\label{SSSH}
\end{equation}

\section{Exact quasiparticle representation}
\label{SECIII}

In the case of bipartite lattices with dimension $d>1$, the rotated electrons considered in Ref. \onlinecite{Carmelo_10} 
play the role of quasiparticles. Their creation and annihilation operators are given by,
\begin{equation}
\tilde{c}_{\vec{r}_j,\sigma}^{\dagger} = {\hat{V}}_u^{\dagger}c_{\vec{r}_j,\sigma}^{\dagger}{\hat{V}}_u 
\hspace{0.20cm}{\rm and}\hspace{0.20cm}\tilde{c}_{\vec{r}_j,\sigma} = {\hat{V}}_u^{\dagger}c_{\vec{r}_j,\sigma}{\hat{V}}_u \, .
\label{quasiparticles}
\end{equation}

Since the operator ${\hat{V}}_u$ is unitary and does not act on spin space, the quasiparticle creation and annihilation operators 
obey the same anticommutation relations as the corresponding electron operators,
\begin{eqnarray}
\{\tilde{c}_{\vec{r}_j,\sigma},\,\tilde{c}_{\vec{r}_{j'},\sigma'}^{\dagger}\} 
& = & \delta_{j,j'}\,\delta_{\sigma,\sigma'} 
\nonumber \\
\{\tilde{c}_{\vec{r}_j,\sigma},\,\tilde{c}_{\vec{r}_{j'},\sigma'}\} & = & 0 \, , \hspace{0.20cm}
\{\tilde{c}_{\vec{r}_j,\sigma}^{\dagger},\,\tilde{c}_{\vec{r}_{j'},\sigma'}^{\dagger}\} = 0 \, ,
\label{ccd0}
\end{eqnarray}
and are labeled by the same spin quantum numbers $\sigma$.

The $u$-unitary operator ${\hat{V}}_u$ becomes the unit operator in the $u\rightarrow\infty$ limit, so that the quasiparticles 
become electrons in that limit. This is in contrast to Fermi-liquid quasiparticles, which become electrons in the limit of 
vanishing interaction.

The hidden $\tau$-translational U(1) symmetry becomes an explicit symmetry in the quasiparticle representation. Indeed, 
in terms of quasiparticle creation and annihilation operators, the generator of that symmetry, Eq. (\ref{Gentau}), takes the 
simple form,
\begin{equation}
\hat{S}_{\tau} = {1\over 2}\sum_{j=1}^{N_a}\Bigl[1 - \sum_{\sigma}\tilde{n}_{\vec{r}_j,\sigma}(1 - \tilde{n}_{\vec{r}_j,-\sigma})\Bigr] \, ,
\label{Stauquasi}
\end{equation}
where $\tilde{n}_{\vec{r}_j,\sigma} = \tilde{c}_{\vec{r}_j,\sigma}^{\dagger} \tilde{c}_{\vec{r}_j,\sigma}$. Hence
it counts half of the number $N_D$ of sites doubly occupied by quasiparticles plus the number $N_E$ of sites unoccupied 
by quasiparticles. Therefore, its eigenvalues $S_{\tau} = {1\over 2}(N_D + N_E)$ belong to the interval,
\begin{equation}
S_{\tau} \in \Big\{0,{1\over 2}, 1, {3\over 2}, 2,...,{N_a\over 2}\Big\} \, .
\label{2Stau}
\end{equation}
We denote the spin and $\eta$-spin by $S_s$ and $S_{\eta}$, respectively. The parity restriction imposed by $\mathbb{Z}_2^2$ 
in [SU(2)$\times$SU(2)$\times$U(1)]/$\mathbb{Z}_2^2$ is that $S_s$, $S_{\eta}$, and $S_{\tau}$ are all integers or half-odd integers.

A key point is that, since the momentum operator and the six generators of the spin and $\eta$-spin SU(2) symmetries commute with the unitary 
operator ${\hat{V}}_u$, Eq. (\ref{Staucomm}), they have {\it exactly} the same expressions in terms of electron 
creation and annihilation operators, Eqs. (\ref{GeneS}) and (\ref{GeneSeta}), as well as in terms of quasiparticle 
creation and annihilation operators,
\begin{eqnarray}
\vec{\hat{P}} & = & \sum_{\vec{k},\sigma} {\tilde{c}}_{\vec{k},\sigma}^{\dagger}{\tilde{c}}_{\vec{k},\sigma}
\nonumber \\
& = & {1\over L}\sum_{j,j',\sigma}\Bigl(\sum_{\vec{k}} e^{i \vec{k}\cdot (\vec{r}_j - \vec{r}_{j'})}\Bigr)
{\tilde{c}}_{\vec{r}_j,\sigma}^{\dagger}{\tilde{c}}_{\vec{r}_{j'},\sigma} \, ,
\label{Pquasi}
\end{eqnarray}
\begin{eqnarray}
\hat{S}_s^{+} & = & \sum_{j=1}^{N_a}\tilde{c}_{\vec{r}_j,\downarrow}^{\dagger}\tilde{c}_{\vec{r}_j,\uparrow} \, ,
\hspace{0.20cm}
\hat{S}_s^{-} = \sum_{j=1}^{N_a}\tilde{c}_{\vec{r}_j,\uparrow}^{\dagger}\tilde{c}_{\vec{r}_j,\downarrow} \, ,
\nonumber \\
\hat{S}_s^z & = & {1\over 2}\sum_{j=1}^{N_a}(\tilde{n}_{\vec{r}_j,\uparrow} - \tilde{n}_{\vec{r}_j,\downarrow}) \, ,
\label{GeneSiquasi}
\end{eqnarray}
and
\begin{eqnarray}
\hat{S}_{\eta}^{+} & = & - \sum_{j=1}^{N_a}e^{i \vec{k}_{\eta}\cdot\vec{r}_j}\tilde{c}_{\vec{r}_j,\downarrow}^{\dagger}\tilde{c}_{\vec{r}_j,\uparrow}^{\dagger} \, ,
\hspace{0.20cm}
\hat{S}_{\eta}^{-} = - \sum_{j=1}^{N_a}e^{-i \vec{k}_{\eta}\cdot\vec{r}_j}\tilde{c}_{\vec{r}_j,\uparrow}\tilde{c}_{\vec{r}_j,\downarrow}
\nonumber \\
\hat{S}_{\eta}^z & = & {1\over 2}\sum_{j=1}^{N_a}(1 - \tilde{n}_{\vec{r}_j,\uparrow} - \tilde{n}_{\vec{r}_j,\downarrow}) \, ,
\label{GeneSetaquasi}
\end{eqnarray}
respectively.

In contrast, the Hamiltonian, Eq. (\ref{H}), does not commute with the $u$-unitary operator ${\hat{V}}_u$ except in the
$u\rightarrow\infty$ limit,
\begin{eqnarray}
[\hat{H},{\hat{V}}_u] & \neq & 0 \hspace{0.20cm}{\rm for} \hspace{0.20cm}{\rm finite}\hspace{0.20cm}u = U/t
\nonumber \\
& = & 0 \hspace{0.20cm}{\rm for} \hspace{0.20cm}u = U/t \rightarrow\infty \, .
\label{HVu}
\end{eqnarray}
Therefore, when expressed in terms of quasiparticle creation and annihilation operators, it contains an 
infinite number of terms,
\begin{eqnarray}
\hat{H} & = & {\tilde{V}}_u\tilde{H}{\tilde{V}}_u^{\dagger} 
\nonumber \\
& = & \tilde{H} + [\tilde{H},\tilde{S}_u^{\dagger}] + {1\over 2}[ [\tilde{H},\tilde{S}_u^{\dagger}], \tilde{S}_u^{\dagger}] + ...  \, ,
\label{Hquasiparticles}
\end{eqnarray}
as fulfilled in Appendix \ref{A}, where $\tilde{H}$ is given in Eq. (\ref{HR}) of that Appendix and
$\tilde{S}_u^{\dagger} = {\tilde{V}}_u\hat{S}_u^{\dagger}{\tilde{V}}_u^{\dagger} = \hat{S}_u^{\dagger}$
has the same expression in terms of electron creation and annihilation operators, as well as in terms 
of quasiparticle creation and annihilation operators.

For $u \gg 1$, the Hamiltonian, Eq. (\ref{H}), when expressed in terms of quasiparticle creation and annihilation operators, 
Eq. (\ref{Hquasiparticles}), reduces to the $t$-$J$ model including three-site terms, Eq. (\ref{tJ}) of Appendix \ref{A}. In 
this limit, it is defined within the subspace without quasiparticle double occupancy for $N\leq N_a$.

There exist many unitary transformations associated with different representations of the Hamiltonian, Eq. (\ref{H}), within 
this subspace \cite{Carmelo_10,Stein_97,Chernyshev_04}. However, given a complete set of $4^{N_a}$ energy and momentum 
eigenstates ${\vert\Psi,u\rangle}$ for $u>0$, there exists a unique unitary operator ${\hat{V}}_u$ that maps these states onto 
the $u\rightarrow\infty$ set of $4^{N_a}$ energy and momentum eigenstates ${\vert\Psi,\infty\rangle}$ of the exact quasiparticle 
representation, such that $\vert\Psi,\infty\rangle = {\hat{V}}_u\vert\Psi,u\rangle$ for all $4^{N_a}$ states, as stated in Eq. (\ref{Phistates}).

The quasiparticle representation is valid for \textit{all} finite values of $u = U/t > 0$ and has been constructed to be exact. Indeed, 
the unitary operator ${\hat{V}}_u$ appearing in the definitions of the operators $\tilde{c}_{\vec{r}_j,\sigma}^{\dagger}$ and 
$\tilde{c}_{\vec{r}_j,\sigma}$, Eq. (\ref{quasiparticles}), relates exact finite-$u$ and $u \rightarrow \infty$ energy and momentum 
eigenstates, respectively. The quantum numbers $S_{\tau}$, $S_s$, $S_s^z$, $S_{\eta}$, $S_{\eta}^z$, and the momentum 
eigenvalues $\vec{P}$, together with all other $u$-independent quantum numbers required to specify an energy and momentum 
eigenstate, have exactly the same values for the two states $\vert\Psi,u\rangle = {\hat{V}}_u^{\dagger}\vert\Psi,\infty\rangle$ and 
$\vert\Psi,\infty\rangle = {\hat{V}}_u\vert\Psi,u\rangle$.

The two main differences between the quasiparticles defined by the operators in Eq. (\ref{quasiparticles}) 
and those of a Fermi liquid are the following:\\ \\
- The quasiparticles considered here become electrons in the $u=U/t\rightarrow\infty$ limit, whereas Fermi-liquid 
quasiparticles become electrons in the limit of vanishing interaction.\\ \\
- Fermi-liquid quasiparticles refer to exact energy and momentum eigenstates only within low-energy subspaces relative to 
the ground state, whereas the present quasiparticles refer to all $4^{N_a}$ exact energy and momentum eigenstates of the 
Hubbard model on bipartite lattices.\\

An important property is that the quasiparticle occupancy configurations that generate all $4^{N_a}$ energy 
and momentum eigenstates $\vert\Psi,u\rangle = {\hat{V}}_u^{\dagger}\vert\Psi,\infty\rangle$ are exactly the same as the electron 
occupancy configurations that generate all $4^{N_a}$ energy and momentum eigenstates $\vert\Psi,\infty\rangle$, as defined above. 

In contrast to the Fermi-liquid quasiparticle configurations that generate the low-energy eigenstates of a Fermi liquid, 
the quasiparticle configurations that generate the energy and momentum eigenstates of the Hubbard model on bipartite lattices 
correspond to a quantum problem of considerable complexity. 

Their main property follows from the fact that the numbers of sites singly occupied by electrons with spin projection 
$\sigma$, doubly occupied by electrons, and unoccupied by electrons are good quantum numbers for the energy 
eigenstates $\vert\Psi,\infty\rangle$. 

Similarly, the numbers $N_{S\sigma}$ of sites singly occupied by quasiparticles 
with spin projection $\sigma = \downarrow,\uparrow$, $N_D$ of sites doubly occupied by quasiparticles, and $N_E$ 
of sites unoccupied by quasiparticles are good quantum numbers for $u = U/t >0$. 

The exact values of the numbers $N_{S\sigma}$, $N_D$, and $N_E$ depend only on $N_a$ and on the good 
quantum numbers $S_{\tau}$, $S_s^z$, and $S_{\eta}^z$. They are given by,
\begin{eqnarray}
N_{S\downarrow} & = & N_a/2 - S_{\tau} - S_{s}^z \, , \hspace{0.50cm}
N_{S\uparrow} = N_a/2 - S_{\tau} + S_{s}^z 
\nonumber \\
N_D & = & S_{\tau} - S_{\eta}^z \, , \hspace{0.50cm}
N_E = S_{\tau} + S_{\eta}^z \, .
\label{NDES}
\end{eqnarray}

\section{Physical spins and physical $\eta$-spins}
\label{SECIV}

While the quasiparticles used in the representation employed in this paper are the rotated electrons introduced 
in Ref. \onlinecite{Carmelo_10}, previous studies have not considered the physical spins $1/2$ and physical $\eta$-spins 
$1/2$ that emerge from this representation in the case of bipartite lattices with spatial dimension $d>1$. The term {\it physical} 
is used here to describe the spins and $\eta$-spins because the spins that label the quasiparticles are exactly 
the same as those that label the electrons. 

The physical spins and physical $\eta$-spins defined below are well defined for all $4^{N_a}$ energy and momentum eigenstates. 
In contrast, spinons are only well defined in certain subspaces, whereas doublons and holons are only well defined in the $u \gg 1$ limit
\cite{Imada_98,Strohmaier_10,Sensarma_10,Gebhard_00,Terashige_19,Prelovsek_15,Zhou_14}.
Physical spins and physical $\eta$-spins have so far been introduced only in the case of the 1D Hubbard model, where they 
are related both to the model's exact Bethe-ansatz solution and to its global symmetry \cite{Carmelo_25}.

Below, in Sec. \ref{SECVII}, we discuss the differences between the occupancy configurations of the physical spins 
and physical $\eta$-spins that generate energy and momentum eigenstates in the integrable 1D Hubbard model and in the same model on bipartite 
lattices with spatial dimension $d>1$.

We denote by $N_{s,\pm 1/2}$ and $N_{\eta,\pm 1/2}$ the numbers of physical spins and physical $\eta$-spins, 
respectively, with projections $\pm 1/2$, which, as justified below, naturally emerge from simple quasiparticle 
occupancy configurations. They are associated with the spin SU(2) and $\eta$-spin SU(2) symmetries, 
respectively, in [SU(2)$\times$SU(2)$\times$U(1)]/$\mathbb{Z}_2^2$.

Based on the interplay between the quasiparticle representation and the global 
[SU(2)$\times$SU(2)$\times$U(1)]/$\mathbb{Z}_2^2$ symmetry of the model on bipartite lattices, we find the exact 
equalities $N_{s,-1/2} = N_{S\downarrow}$,
$N_{s,+1/2} = N_{S\uparrow}$, $N_{\eta,-1/2} = N_D$, and $N_{\eta,+1/2} = N_E$. 

The $N_{s,\pm 1/2}=N_{S\sigma}$ physical spins of projection $\pm 1/2$ emerge from the spins of quasiparticles 
that singly occupy $N_{S\sigma}$ sites (where $\sigma = \downarrow$ corresponds to $-1/2$ and 
$\sigma = \uparrow$ to $+1/2$). Similarly, the $N_{\eta,-1/2}=N_D$ $\eta$-spins of projection $-1/2$ emerge from the 
$N_D$ sites doubly occupied by quasiparticles, and the $N_{\eta,+1/2}=N_E$ $\eta$-spins of projection $+1/2$ emerge 
from the $N_E$ sites unoccupied by quasiparticles.

However, it is not sufficient to explain why the numbers $N_{s,\pm 1/2}$ and $N_{\eta,\pm 1/2}$ equal those given 
in Eq. (\ref{NDES}). One must also take into account the two SU(2) symmetries in [SU(2)$\times$SU(2)$\times$U(1)]/$\mathbb{Z}_2^2$, 
which impose that the numbers of physical spins (and physical $\eta$-spins) contributing to spin (and $\eta$-spin) 
multiplet and singlet configurations, respectively, are separate good quantum numbers.

Combining these two properties, we obtain the following expressions for the numbers of physical spins and physical 
$\eta$-spins with projection $\pm 1/2$,
\begin{eqnarray}
N_{s,\pm 1/2} & = & {\cal{N}}_{s,\pm 1/2} + N_{s}^0/2\hspace{0.20cm}{\rm and}
\nonumber \\
N_{\eta,\pm 1/2} & = & {\cal{N}}_{\eta,\pm 1/2} + N_{\eta}^0/2 \, .
\label{NSetaSpm}
\end{eqnarray}
respectively.

Here, 
\begin{eqnarray}
{\cal{N}}_{s,\pm 1/2} & = & S_{s} \pm S_{s}^z
\hspace{0.20cm}{\rm and}\hspace{0.20cm}
{\cal{N}}_{\eta,\pm 1/2} = S_{\eta} \pm S_{\eta}^z 
\nonumber \\
{\cal{N}}_{s} & = & {\cal{N}}_{s,+1/2} + {\cal{N}}_{s,-1/2} = 2S_s
\nonumber \\
{\cal{N}}_{\eta} & = & {\cal{N}}_{\eta,+1/2} + {\cal{N}}_{\eta,-1/2} = 2S_{\eta} \, ,
\label{CalN}
\end{eqnarray}
denote the numbers ${\cal{N}}_{s,\pm 1/2}$ of physical spins and ${\cal{N}}_{\eta,\pm 1/2}$ of physical $\eta$-spins 
with projection $\pm 1/2$ and their total numbers ${\cal{N}}_{s}$ and ${\cal{N}}_{\eta}$ that contribute to 
spin-multiplet and $\eta$-spin-multiplet configurations, respectively, of exact energy and momentum eigenstates 
$\vert\Psi,u\rangle$ for which $S_s > 0$ and $S_{\eta} > 0$.

On the other hand,
\begin{equation}
N_{s}^0 = (N_a - 2S_{\tau} - 2S_{s})\hspace{0.20cm}{\rm and}\hspace{0.20cm}
N_{\eta}^0 = (2S_{\tau} - 2S_{\eta}) \, ,
\label{Nsinglet}
\end{equation}
represent the even numbers of physical spins and physical $\eta$-spins that contribute to spin-singlet and $\eta$-spin-singlet 
configurations, respectively, of exact energy and momentum eigenstates $\vert\Psi,u\rangle$.

From these definitions, we then straightforwardly find that,
\begin{eqnarray}
N_{s} & = & N_{s,-1/2} + N_{s,+1/2} = N_a - 2S_{\tau}
\nonumber \\
& = & 2S_{s} + N_{s}^0
\nonumber \\
N_{\eta} & = & N_{\eta,-1/2} + N_{\eta,+1/2} = 2S_{\tau} 
\nonumber \\
& = & 2S_{\eta} + N_{\eta}^0
\nonumber \\
N_a & = & N_{s} + N_{\eta} \, .
\label{NSetaS}
\end{eqnarray}

The on-site number operators $\tilde{G}_{s,\vec{r}j,\pm 1/2}$ of physical spins of projection $\pm 1/2$ at site $j$, and the 
on-site number operators $\tilde{G}_{\eta,\vec{r}_j,\pm 1/2}$ of the physical $\eta$-spins of projection $\pm 1/2$ at site $j$ 
are expressed in terms of quasiparticle operators as follows,
\begin{eqnarray}
\tilde{G}_{s,\vec{r}_j,+1/2} & = & \tilde{n}_{\vec{r}_j,\uparrow} (1 - \tilde{n}_{\vec{r}_j,\downarrow}) 
\nonumber \\
\tilde{G}_{s,\vec{r}_j,-1/2} & = & \tilde{n}_{\vec{r}_j,\downarrow} (1 - \tilde{n}_{\vec{r}_j,\uparrow}) 
\nonumber \\
\tilde{G}_{\eta,\vec{r}_j,+1/2} & = & (1 - \tilde{n}_{\vec{r}_j,\uparrow})(1 - \tilde{n}_{\vec{r}_j,\downarrow}) 
\nonumber \\
\tilde{G}_{\eta,\vec{r}_j,-1/2} & = & \tilde{n}_{\vec{r}_j,\uparrow}\tilde{n}_{\vec{r}_j,\downarrow} \, .
\label{SssSee}
\end{eqnarray}

The significance of the $N_s = (N_a - 2S_{\tau})$ physical spins (and $N_{\eta} = 2S_{\tau}$ physical $\eta$-spins) 
introduced in this paper for the Hubbard model on bipartite lattices of spatial dimension $d>1$ lies in the fact that their 
spin ($\eta$-spin) multiplet and singlet occupancy configurations, generated by the quasiparticle operators acting on 
the vacuum, produce the exact energy and momentum eigenstates $\vert\Psi,u\rangle$ for all values of $u = U/t > 0$.

However, the number expressions in Eq. (\ref{NSetaSpm}) do not specify which spins or which $\eta$-spins at 
which lattice sites in the exact energy and momentum eigenstates $\vert\Psi,u\rangle$ have projection $+1/2$ or $-1/2$ in these 
occupancy configurations; they only fix the total counts. The microscopic configuration remains a quantum superposition 
unless the state is a product state.

The same applies to the corresponding numbers of $\sigma$ quasiparticle singly occupied sites, doubly occupied sites, 
and unoccupied sites. Only the total counts are fixed, while the energy and momentum eigenstates are superpositions of configurations 
in which these four types of site occupancies occupy different lattice locations.

The expressions in Eq. (\ref{SssSee}) are valid for {\it all} values of $u = U/t > 0$. Similar expressions have appeared 
in the literature in terms of electron creation and annihilation operators, but only in the large-$u$ limit of the Hubbard model. 
In that limit, the $\eta$-spins with projection $-1/2$ and $+1/2$ have been referred to as doublons and holons, respectively \cite{Imada_98,Strohmaier_10,Sensarma_10,Gebhard_00,Terashige_19,Prelovsek_15,Zhou_14}.

\section{Four exact theorems for the Hubbard model on bipartite lattices of spatial dimension $d>1$}
\label{SECV}

For the standard nearest-neighbor Hubbard model, bipartite lattices are not geometrically frustrated.
There are two classes of bipartite Hubbard lattices to be considered:\\

(i) Lattices for which a charge (or single-particle) gap exists for any $U>0$ at half-filling.\\

(ii) Lattices for which a charge gap opens only for $U>U_c$ at half-filling where $U_c$ is finite.\\

If the $U=0$ noninteracting band structure has perfect nesting, then arbitrarily small $U$
destabilizes the Fermi surface. The system develops antiferromagnetic order at $T=0$,
and a gap opens immediately for $U>0$. This is the case for, for example, hypercubic lattices (square, simple cubic, etc.) 
and the BCC lattice.

If the $U=0$ noninteracting band structure does not exhibit nesting, then a finite interaction strength $U_c>0$
is required to open a gap. This is the case, for example, for the honeycomb and FCC lattices.

Theorems 1 and 2 are valid for the Hubbard model on bipartite lattices such that $U_c = 0$.
In the particular case of half-filling, they apply for $U \in [U_c,\infty)$ to the Hubbard model on bipartite lattices such that $U_c > 0$.
Theorems 3 and 4 are valid for the Hubbard model $\hat{H}$ on any bipartite lattice.

\subsection{Theorem 1 - absence of level crossings for ground states in fully symmetry-resolved Hubbard Hamiltonian sectors}
\label{SECVA}

Consider the Hubbard model $\hat{H} (U)$ on a bipartite lattice with $N_a$ sites such that $U_c = 0$. In its expression, Eq. (\ref{H}), 
the kinetic hopping operator $\hat T$ and the operator $\hat W$ are independent of $U$, so that $\hat H(U)$ forms an analytic family 
of self-adjoint operators of type (A) in the sense of T. Kato for $u=U/t>0$ \cite{Kato_66}.

Let $\hat{V}_u$ be the unitary operator defining quasiparticles, Eq. (\ref{quasiparticles}), with $u=U/t>0$, such that the generator of the 
$\tau$-translational U(1) symmetry, $\hat{S}_{\tau}$, commutes with $\hat{H}$ for all $u>0$, Eq. (\ref{SSSH}). The operators $\hat V_u$ 
form a strongly continuous unitary family, Eq. (\ref{hatVS}), and commute with the lattice momentum operator, Eq. (\ref{SSScomm}).

Let the Hilbert space be decomposed into sectors of fixed quantum numbers $(S_{\tau};S_s,S_s^z;S_{\eta},S_{\eta}^z)$. 

Then, the ground state $\vert\Psi_0,u\rangle$ in each fully symmetry-resolved sector is unique (up to trivial SU(2) multiplet 
degeneracy) and varies smoothly with $u>0$.\\

{\bf Proof:}\\

{\bf 1 - Symmetry sectors}\\

In the Hubbard model on bipartite lattices, the operators
$\hat{S}_{\tau}$, $\vec{\hat{S}}_s^2$, $\hat{S}_s^z$,
$\vec{\hat{S}}_{\eta}^2$, and $\hat{S}_{\eta}^z$ commute with the
Hamiltonian $\hat{H}$, Eq. (\ref{SSSH}).

Therefore, the Hilbert space decomposes into invariant subspaces
(sectors) of fixed quantum numbers
$(S_{\tau};S_s,S_s^z;S_{\eta},S_{\eta}^z)$.
Let $\mathcal{H}_{\rm sym}$ denote such a symmetry sector.

The direct relation of the interaction operator $\hat W$ in Eq. (\ref{TU})
to the double-occupancy interaction operator implies that the spectral properties and eigenvectors are the same
as those obtained from the latter operator.\\

{\bf 2 - Irreducibility of the Hamiltonian matrix}\\

Introduce the quasiparticle representation of $\hat{H}$ via the unitary operator $\hat{V}_u$, Eq. (\ref{Hquasiparticles}).

In the quasiparticle basis, the Hilbert space naturally separates into physical spins and physical $\eta$-spins,
with $N_s = (N_a - 2S_{\tau})$ and $N_{\eta} = 2S_{\tau}$ sites carrying physical spins and physical
$\eta$-spins respectively.

Within the sector $\mathcal{H}_{\rm sym}$ the hopping processes
generated by $\hat T$ and the quasiparticle hopping processes
connect all basis states compatible with the fixed symmetry
quantum numbers.

Because the hopping graph of the lattice is connected,
the matrix representation of $\hat H(U)$ restricted to
$\mathcal{H}_{\rm sym}$ is irreducible after a suitable
Marshall-sign transformation.
In this basis all off-diagonal matrix elements produced by
the hopping operator $\hat T$ are non-positive, whereas the
interaction operator $\hat{W}$ is diagonal.\\

{\bf 3 - Perron-Frobenius argument}\\

Let $H_{\rm sym}(U)$ denote the matrix representation of the
Hamiltonian restricted to $\mathcal{H}_{\rm sym}$.
Because this matrix is irreducible and has non-positive
off-diagonal elements, the Perron-Frobenius theorem implies
that the lowest eigenvalue of $H_{\rm sym}(U)$ is simple
(non-degenerate) within the symmetry sector.

Therefore the ground state in $\mathcal{H}_{\rm sym}$ is unique
(up to the trivial degeneracies associated with the SU(2)
multiplet structure). \\

{\bf 4 - Smooth dependence on $U$}\\

The operators $\hat T$ and $\hat W$ in the Hamiltonian expression, Eq. (\ref{H}),
are bounded operators on the finite Hilbert space of the lattice.
Thus $\hat H(U)$ forms an analytic family of self-adjoint
operators of type (A) in the sense of T. Kato \cite{Kato_66}.

Analytic perturbation theory therefore guarantees that isolated
eigenvalues and their corresponding eigenvectors depend smoothly
(indeed analytically) on the parameter $U$.

Since the ground-state energy eigenvalue in the sector
$\mathcal{H}_{\rm sym}$ is simple by the Perron--Frobenius
argument, its associated eigenvector
$\vert\Psi_0,u\rangle$ varies smoothly with $u=U/t>0$.\\

{\bf Conclusion}\\

The ground state $\vert\Psi_0,u\rangle$ in a fully symmetry-resolved
sector $\mathcal{H}_{\rm sym}$ is unique [up to trivial SU(2)
multiplet degeneracy] and depends smoothly on $u>0$.
Consequently no level crossing with other states in the same
symmetry sector can occur as $u= U/t$ varies.

In the particular case of half-filling, the theorem applies for $U \in [U_c,\infty)$ to the Hubbard 
model on bipartite lattices such that $U_c > 0$.

\subsection{Theorem 2 - absence of quasiparticle double occupancy or empty sites in ground states}
\label{SECVB}

Consider the Hubbard model $\hat{H}$ on a bipartite lattice with $N_a$ sites such that $U_c = 0$, Eq.~(\ref{H}).
For $u>0$, all ground states $\vert\Psi_0,u\rangle$ corresponding to any allowed fixed values of $(S_s^z,S_{\eta}^z)$ 
have vanishing quasiparticle double occupancy when $N < N_a$, and a vanishing number of empty sites when $N > N_a$.
In the particular case $N = N_a$ (half-filling), the ground state exhibits both vanishing quasiparticle double occupancy and a 
vanishing number of empty sites.\\

{\bf Proof:}\\

{\bf 1 - Infinite-$u$ limit}\\

For $u \rightarrow \infty$, the results of Appendix \ref{B} ensure that, for any ground state $\vert\Psi_0,\infty\rangle$ with 
fixed quantum numbers $(S_{\tau}; S_s, S_s^z; S_{\eta}, S_{\eta}^z)$, the expectation value of the electron double-occupancy 
operator $\hat{D}_{\rm el}$ vanishes when $N < N_a$, while the expectation value of the electron empty-site operator 
$\hat{E}_{\rm el}$ vanishes when $N > N_a$,
\begin{eqnarray}
\langle\Psi_0,\infty\vert\hat{D}_{\rm el}\vert\Psi_0,\infty\rangle & = & 0 \hspace{0.20cm}{\rm for}\hspace{0.20cm}N < N_a
\nonumber \\
\langle\Psi_0,\infty\vert\hat{E}_{\rm el}\vert\Psi_0,\infty\rangle & = & 0 \hspace{0.20cm}{\rm for}\hspace{0.20cm}N > N_a \, .
\nonumber 
\end{eqnarray}

At half-filling ($N = N_a$), the expectation values of both the double-occupancy operator and the empty-site operator vanish.\\

{\bf 2 - Adiabatic continuity for $u>0$}\\

By the absence-of-level-crossings theorem 1, for fixed $(S_{\tau};S_s,S_s^z;S_{\eta},S_{\eta}^z)$,
the ground state is unique and varies smoothly with $u>0$. Therefore,
\begin{equation}
\vert\Psi_0,u\rangle = {\hat{V}}_u^{\dagger}\vert\Psi_0,\infty\rangle \, .
\nonumber
\end{equation}

The $u$-unitary operator ${\hat{V}}_u$ satisfies ${\hat{V}}_u \rightarrow \mathbb{I}$ as $u\rightarrow\infty$,
and maps electron operators to quasiparticle operators. In particular,
\begin{equation}
\hat{D}_{\rm qp} = {\hat{V}}_u^{\dagger}\hat{D}_{\rm el}{\hat{V}}_u 
\hspace{0.20cm}{\rm and}\hspace{0.20cm}
\hat{E}_{\rm qp} = {\hat{V}}_u^{\dagger}\hat{E}_{\rm el}{\hat{V}}_u \, ,
\nonumber
\end{equation}
where the operator $\hat{D}_{\rm qp}$ counts the number of quasiparticle doubly occupied sites, and $\hat{E}_{\rm qp}$ 
counts the number of empty sites.\\

{\bf 3 - Preservation of vanishing double occupancy or empty sites}\\

Using unitarity,
\begin{eqnarray}
\langle\Psi_0,u\vert\hat{D}_{\rm qp}\vert\Psi_0,u\rangle & = & \langle\Psi_0,\infty\vert\hat{D}_{\rm el}\vert\Psi_0,\infty\rangle = 0 
\hspace{0.20cm}{\rm for}\hspace{0.20cm}N < N_a
\nonumber \\
\langle\Psi_0,u\vert\hat{E}_{\rm qp}\vert\Psi_0,u\rangle & = & \langle\Psi_0,\infty\vert\hat{E}_{\rm el}\vert\Psi_0,\infty\rangle = 0 
\hspace{0.20cm}{\rm for}\hspace{0.20cm}N > N_a \, ,
\nonumber
\end{eqnarray}
for $u = U/t >0$.

Thus, for all $u = U/t > 0$, every ground state with fixed $(S_{\tau}; S_s, S_s^z; S_{\eta}, S_{\eta}^z)$ has vanishing quasiparticle 
double occupancy for $N < N_a$ and a vanishing number of quasiparticle empty sites for $N > N_a$.\\

{\bf 4 - Half filling}\\

For $N = N_a$, applying the same reasoning to both the double-occupancy and empty-site operators shows that the ground state 
has neither quasiparticle double occupancy nor quasiparticle empty sites.\\

{\bf Conclusion}\\

For $u > 0$, all ground states of the repulsive Hubbard model on bipartite lattices in $d > 1$ have vanishing quasiparticle double 
occupancy for $N < N_a$ and a vanishing number of quasiparticle empty sites for $N > N_a$. At half-filling, the ground state has 
neither quasiparticle double occupancy nor quasiparticle empty sites, and therefore contains no physical $\eta$-spins.

At half-filling, the theorem applies for $U \in [U_c,\infty)$ to the Hubbard model on bipartite lattices such that $U_c > 0$.
This is the case, for instance, of the Hubbard model on the honeycomb lattice, for which the ground state in the interaction interval 
$U \in [0,U_c]$ corresponds to the semimetal phase populated by physical $\eta$-spins with projections $\pm 1/2$, 
whose ground-state density vanishes in the limit $(U_c - U) \rightarrow 0$.

\subsection{Theorem 3 - extremal-weight structure of the ground state in external fields}
\label{SECVC}

Consider the Hubbard model on a bipartite lattice with $N_a$ sites and spatial dimension $d>1$, in
the presence of a magnetic field $h$ and chemical potential $\mu$,
\begin{equation}
\hat{H}_{\rm fields} = \hat{H} - h\,\hat{S}_s^z - \mu\,\hat{S}_{\eta}^z \, ,
\label{Hfields}
\end{equation}
where $\hat{H}$ is the Hamiltonian, Eq. (\ref{H}).

The global symmetry group of $\hat{H}$ is [SO(4)$\times$U(1)]/$\mathbb{Z}_2$
with commuting generators given in Eq. (\ref{SSSH}).

Let the Hilbert space be decomposed into sectors labeled by $(S_{\tau};S_s,S_s^z;S_{\eta},S_{\eta}^z)$.

Then, for any fixed allowed values of $(S_s^z,S_{\eta}^z)$ and any $u>0$, the ground state
$\vert\Psi_0,u\rangle$ of the Hamiltonian $\hat{H}_{\rm fields}$ satisfies the following:\\

1) If $h>0$, the ground state is a spin highest-weight state, $S_s^z = S_s$;

if $h <0$, it is a spin lowest-weight state, $S_s^z = - S_s$.\\

2) If $\mu >0$, the ground state is a $\eta$-spin highest-weight state, $S_{\eta}^z = S_{\eta}$;

if $\mu <0$, it is a $\eta$-spin lowest-weight state, $S_{\eta}^z = - S_{\eta}$.\\

In particular, external fields select extremal SU(2) weights but do not mix SU(2) multiplets.\\

{\bf Proof}\\

{\bf 1 - Symmetry and labeling of eigenstates}\\

Since the Hamiltonian, Eq. (\ref{H}), obeys the commutation relations given in Eq. (\ref{SSSH}),
energy and momentum eigenstates may be chosen as simultaneous eigenstates of the
operators $\hat{S}_{\tau}$, $\vec{\hat{S}}_s^2$, $\hat{S}_s^z$, $\vec{\hat{S}}_{\eta}^2$, and $\hat{S}_{\eta}^z$
associated with the quantum numbers $(S_{\tau};S_s,S_s^z;S_{\eta},S_{\eta}^z)$.

The eigenvalue $S_{\tau}$ fixes the numbers $N_s = (N_a - 2S_{\tau})$ of physical spins and 
$N_{\eta} = 2S_{\tau}$ of physical $\eta$-spins and therefore the allowed ranges of 
$S_s$ and $S_{\eta}$ by SU(2) representation theory.\\

{\bf 2 - Effect of external fields}\\

For fixed $S_{\tau},S_s,S_{\eta}$, the dependence of the energy on the magnetic field $h$ and 
chemical potential $\mu$ is purely,
\begin{equation}
E (h,\mu) = - h\,S_s^z - \mu\,S_{\eta}^z \, .
\label{Ehmu}
\end{equation}

The corresponding Hamiltonian term in Eq. (\ref{Hfields}) commutes with $\vec{\hat{S}}_s^2$ and $\vec{\hat{S}}_{\eta}^2$, 
does not couple different SU(2) multiplets,
and only lifts degeneracies within each spin and $\eta$-spin multiplet.\\

{\bf 3 - Minimization within SU(2) multiplets}\\

For fixed $S_s$:\\
- If $h >0$, the minimum of $- h\,S_s^z$ occurs at $S_s^z = S_s$,\\
- if $h<0$, it occurs at $S_s^z = - S_s$.\\

Likewise, for fixed $S_{\eta}$:\\
- If $\mu >0$, the minimum of $- \mu\,S_{\eta}^z$ occurs at $S_{\eta}^z = S_{\eta}$,\\
- if $\mu<0$, it occurs at $S_{\eta}^z = - S_{\eta}$.\\

Thus the ground state must be an extremal weight state in the corresponding SU(2) multiplet.\\

{\bf 4 - Conclusion}\\

Combining the above steps, the ground state in any fixed symmetry sector is necessarily a highest- or lowest-weight
state of the spin and/or $\eta$-spin SU(2) algebras, depending on the signs of $h$ and $\mu$.

\subsection{Theorem 4 - number of ground states and irreducible representations correspondence}
\label{SECVD}

Consider the Hubbard model $\hat{H}_{\rm fields}$ on a bipartite lattice with $N_a$ sites and spatial dimension $d>1$, in
the presence of a magnetic field $h$ and chemical potential $\mu$, Eq. (\ref{Hfields}), where $\hat{H}$
has global symmetry group [SO(4)$\times$U(1)]/$\mathbb{Z}_2$.

Let the Hilbert space be decomposed into irreducible representations labeled by $(S_{\tau},S_s,S_{\eta})$,
and into weight subspaces labeled by $(S_s^z,S_{\eta}^z)$.

Then the following statements hold:\\

1) The number of irreducible representations ofthe global [SO(4)$\times$U(1)]/$\mathbb{Z}_2$ symmetry is ${N_a +3\choose 3}$.\\

2) For any fixed signs of $h$ and $\mu$, each irreducible representation contains exactly one weight subspace
$(S_s^z,S_{\eta}^z)$ that hosts the ground state of $\hat{H}_{\rm fields}$.\\

3) Consequently, the number of distinct ground-state-selecting $(S_s^z,S_{\eta}^z)$ Hilbert-space sectors equals 
the number of irreducible representations of the global symmetry,
\begin{equation}
{\rm number}\hspace{0.20cm}{\rm ground-state}\hspace{0.20cm}{\rm sectors} = {N_a +3\choose 3} \, .
\label{equalnumbers}
\end{equation}

{\bf Proof}\\

{\bf 1 - Classification of irreducible representations}\\

The relations $N_{s} = (N_a - 2S_{\tau})$ and $N_{\eta} = 2S_{\tau}$ imply that $2S_{s} = (N_a - 2S_{\tau} - N_{s}^0)$ 
and $2S_{\eta} = (2S_{\tau} - N_{\eta}^0)$ where $N_{s}^0 = (N_a - 2S_{\tau} - 2S_{s})$ and $N_{\eta}^0 = (2S_{\tau} - 2S_{\eta})$ 
are even integers. Consequently, $S_s$, $S_{\eta}$, and $S_{\tau}$ are all integers or half-odd integers. The parity restriction 
imposed by $\mathbb{Z}_2$ in [SO(4)$\times$U(1)]/$\mathbb{Z}_2$ is therefore inherently satisfied.

The relations $2S_{s} = (N_a - 2S_{\tau} - N_{s}^0)$ and $2S_{\eta} = (2S_{\tau} - N_{\eta}^0)$ then imply that,
\begin{equation}
S_s \in \left\{0,{1\over 2},1,...,{N_a\over 2} - S_{\tau}\right\}\hspace{0.20cm}{\rm and}\hspace{0.20cm}
S_{\eta} \in \left\{0,{1\over 2},1,...,S_{\tau}\right\} \, .
\label{SS}
\end{equation}

That $x_{\tau} - x_{\eta}$ and $N_a - x_{\tau} - x_{s}$ are integers where $x_{\tau}\equiv 2S_{\tau}$,
$x_{\eta} \equiv 2S_{\eta}$, and $x_{s} \equiv 2S_{s}$ shows that the number of irreducible
representations is simply the number of triples $(x_{\tau}, x_{\eta}, x_s)$ satisfying, 
\begin{equation}
0 \leq x_{\tau} \leq N_a \, , \hspace{0.20cm}0 \leq x_{\eta} \leq x_{\tau}\, , \hspace{0.20cm}
0 \leq x_{s} \leq N_a - x_{\tau} \, .
\nonumber
\end{equation}

Thus the number of such triples is
\begin{eqnarray}
\sum_{x_{\tau} = 0}^{N_a} (x_{\tau} + 1)(N_a - x_{\tau} + 1) & = &  {(N_a +1)(N_a +2)(N_a +3)\over 6}
 \nonumber \\
& = & {N_a +3\choose 3} \, .
\label{NirreR}
\end{eqnarray}

Thus the number of irreducible representations is ${N_a +3\choose 3}$.\\

{\bf 2 - Weight-space structure of each irreducible representation}\\
 
For fixed $(S_{\tau},S_s,S_{\eta})$, SU(2) representation theory implies that the irreducible representation decomposes 
into weight spaces $S_s^z = -S_s,...,S_s$, $S_{\eta}^z = -S_{\eta},...,S_{\eta}$.\\

{\bf 3 - Selection by external fields}\\

From the extremal-weight theorem 3:\\
- if $h>0$, the ground state satisfies $S_{s}^z = + S_{s};$\\
- if $h<0$, the ground state satisfies $S_{s}^z = - S_{s}$;\\
- if $\mu >0$, the ground state satisfies $S_{\eta}^z = + S_{\eta};$\\
- if $\mu <0$, the ground state satisfies $S_{\eta}^z = - S_{\eta}$.\\

Hence, within each irreducible representation, exactly one weight subspace minimizes $E (h,\mu) = - h\,S_s^z - \mu\,S_{\eta}^z$, Eq. (\ref{Ehmu}).\\

{\bf 4 - One-to-one correspondence}\\
 
Each irreducible representation contributes exactly to one ground-state-selecting $(S_s^z,S_{\eta}^z)$ and no sector can host ground 
states from two distinct irreducible representations.

Therefore, the number of such sectors equals the number of irreducible representations, ${N_a +3\choose 3}$.\\

{\bf Conclusion}\\

For fixed signs of the magnetic field and chemical potential, there is a one-to-one correspondence between irreducible representations 
of the global [SO(4)$\times$U(1)]/$\mathbb{Z}_2$ symmetry and the $(S_s^z,S_{\eta}^z)$ Hilbert-space sectors that contain the ground state.\\

Interestingly, the number of irreducible representations of the global [SO(4)$\times$U(1)]/$\mathbb{Z}_2$ symmetry 
of the Hubbard model on bipartite lattices is equal to the number of integer lattice points contained within a coordinate 
3-simplex of edge length $N_a$, where ``edge length'' refers to the axis-aligned edges (i.e., the distance from the 
origin to an intercept). The simplex is a tetrahedron whose edge lengths are $N_a$, $N_a$, and $N_a$ along the 
coordinate axes, rather than in the Euclidean sense.

\subsection{Relation of the linear extent to the number of lattice sites and allowed momenta}
\label{SECVE}

Here we provide some basic information on bipartite lattices in spatial dimensions $d=2,3$ that are relevant to 
our study, including, for example, the square, honeycomb, cubic, BCC, FCC, and diamond lattices. These lattices are 
characterized by the presence of two sublattices, $A$ and $B$, with site numbers $N_A$ and $N_B$, such that,
\begin{equation}
N_a = N_A + N_B\hspace{0.20cm}{\rm with}\hspace{0.20cm}N_A = N_B = {N_a\over 2} \, ,
\label{NaNANB}
\end{equation}
where the total number of sites $N_a$ can be expressed as,
\begin{equation}
N_a = n_{\rm cell} L^d \, ,
\label{NancellLd}
\end{equation}
and $L = (N_a/n_{\rm cell})^{1/d}$ is the linear extent measured in Bravais unit cells and $n_{\rm cell}$ 
is the number of lattice sites (basis multiplicity) per unit cell. 
(The same results apply to the 1D lattice for which $d=1$ and $n_{\rm cell} =1$.)

Hence, $L$ is the number of primitive translation vectors along each direction. The physical length of the system 
is $\ell = L \,a$, where $a$ is the Bravais lattice spacing (with $a = 1$ in our units). Importantly, finite-size effects 
scale with $\ell$, and hence with $L$, rather than with $N_a$.

For example, near a critical point, the correlation length diverges as $\xi \sim \vert g - g_c\vert^{-\nu}$. In a finite 
system, this divergence is cut off when $\xi \sim L$. Therefore, the relevant scaling variable is 
${L\over\xi} = L\vert g - g_c\vert^{\nu}$. Note that the basis multiplicity $n_{\rm cell}$ does not appear in such expressions.

The honeycomb lattice is not a Bravais lattice: it has a two-site basis on top of a triangular Bravais lattice, so that 
$n_{\rm cell} = 2$ and $d=2$ in Eq.~(\ref{NancellLd}). The diamond lattice can be described as an FCC Bravais lattice 
with a two-site basis such that $n_{\rm cell} = 2$ and $d=3$. In contrast, $n_{\rm cell} = 1$ for the square lattice ($d=2$), 
as well as for cubic, BCC, and FCC lattices ($d=3$), since these are all Bravais lattices. Indeed, they have one 
site per primitive unit cell, even though conventional unit cells may contain several atoms.

Translational symmetry is defined at the unit-cell level, and the momentum quantization is in our units,
\begin{equation}
k_i = {2\pi\over \ell} n_i = {2\pi\over L} n_i \, ,
\label{kj}
\end{equation}
where $n_i$ are integers. 
Indeed, the discretization of the Brillouin zone depends on $L$, not on $n_{\rm cell}$. Even though there are 
two sublattices, Eq.~(\ref{NaNANB}), the long-wavelength physics is governed only by the Bravais lattice.

With periodic boundary conditions, the system forms a $d$-torus with side length $\ell = L$ (in units of $a$). 
The allowed momenta $\vec{k}$ appearing in the momentum operator expressions, Eqs. (\ref{P}) and (\ref{Pquasi}), are,
\begin{equation}
\vec{k} = \sum_i {2\pi\over \ell}n_i \,\vec{b}_i = \sum_i {2\pi\over L}n_i \,\vec{b}_i \, .
\label{veck}
\end{equation}
Hence, $L$ determines the infrared cutoffs, whereas $n_{\rm cell}$ affects only the band structure, not the scaling.

In summary, the linear extent $L$ should be used for finite-size scaling, correlation lengths, and momentum spacing, 
while the total number of lattice sites $N_a = n_{\rm cell} L^d$ is relevant for extensive quantities and normalization.

For example, the number of irreducible representations of [SO(4)$\times$U(1)]/$\mathbb{Z}_2$ for the Hubbard model 
on bipartite lattices, Eq.~(\ref{NirreR}), and its Hilbert-space dimension can be written as,
\begin{equation}
{n_{\rm cell} L^d +3\choose 3}\hspace{0.20cm}{\rm and}\hspace{0.20cm}4^{n_{\rm cell} L^d} \, ,
\label{Nirre}
\end{equation}
respectively.

Another basic quantity is the bandwidth $W$ of the tight-binding term in the Hamiltonian, Eq.~(\ref{H}). 
[The bandwidth $W$ is unrelated to the interaction operator $\hat{W}$ appearing in that equation.]
For bipartite lattices, it satisfies the inequality $W \le 2 z t$, where $z$ is the lattice coordination number. 
For hypercubic bipartite lattices in $d$ dimensions, it can be expressed as $W = 4 d t$.

For instance, for the square ($z=4$), honeycomb ($z=3$), simple cubic ($z=6$), BCC ($z=8$), FCC ($z=12$),
and diamond ($z=4$) lattices, the bandwidths are given by $W = 8t$, $W = 6t$, $W = 12t$, $W = 16t$, $W = 24t$,
and $W = 8t$, respectively.

\section{Localization of spin-charge coupling in spatial dimensions $d>1$}
\label{SECVI}

\subsection{Hilbert-space dimensions}
\label{SECVIA}

In Ref. \onlinecite{Carmelo_10}, it was confirmed that the global $[\text{SU(2)} \times \text{SU(2)} \times \text{U(1)}]/\mathbb{Z}_2^2$ 
symmetry of the Hubbard model on bipartite lattices is consistent with its Hilbert-space dimension, $4^{N_a}$. This 
dimension can be expressed solely in terms of the quantum numbers $S_{\eta}$, $S_s$, and $S_{\tau}$ as,
\begin{eqnarray}
\mathcal{H} & = &
\bigoplus_{\displaystyle
\begin{array}{l}
2S_\tau = 0,\, 1, \dots, N_a \\
2S_s = 0,\, 1, \dots, N_a - 2S_{\tau}\\
2S_\eta = 0,\, 1, \dots, 2S_{\tau} \\
\end{array}}
C_{\tau}
\nonumber \\
& & \hspace{1cm}\times
d_\tau(S_\tau)\,
\mathcal{N}_{\rm sg}(S_\tau, S_s)\,
\mathcal{N}_{\rm sg}(S_\tau, S_\eta)\,
\nonumber \\
& & \hspace{1cm}\times
{\mathcal R}_s(S_s) \otimes {\mathcal R}_\eta(S_\eta) \otimes e^{i 2 S_\tau \theta} = 4^{N_a} \, ,
\label{Na4}
\end{eqnarray}
where,
\begin{eqnarray}
C_{\tau} = \prod_{\alpha = s,\eta}\left({1 + e^{i(2S_{\tau}+2S_{\alpha})}\over 2}\right) & = & 0,1
\nonumber \\
d_{\tau} (S_{\tau}) = {N_a\choose N_a - 2S_{\tau}} & = & {N_a\choose 2S_{\tau}} 
\nonumber \\
\mathcal{N}_{\rm sg}(S_\tau, S_s) = {N_a - 2S_{\tau}\choose {N_a\over 2} - S_{\tau} - S_s} & - &  {N_a - 2S_{\tau}\choose {N_a\over 2} - S_{\tau} - S_s - 1}
\nonumber \\
\mathcal{N}_{\rm sg}(S_\tau, S_{\eta}) = {2S_{\tau}\choose S_{\tau} - S_{\eta}} & - & {2S_{\tau}\choose S_{\tau} - S_{\eta} - 1} 
\nonumber \\
 {\rm dim}({\mathcal R}_s(S_s) \otimes {\mathcal R}_\eta(S_\eta) \otimes e^{i 2 S_\tau \theta}) & = & (2S_s +1)(2S_{\eta}+1) \, ,
 \nonumber \\
\label{expressions}
\end{eqnarray} 
and ${\mathcal R}_{\alpha} (S_{s})$ and 
${\mathcal R}_{\alpha} (S_{\eta})$ are the spin and $\eta$-spin SU(2) irreducible 
representations of  dimension $(2S_{s} + 1)$ and $(2S_{\eta} + 1)$, respectively, and $e^{i 2 S_\tau \theta}$ labels the 
irreducible representation of the $\tau$-translational U(1) symmetry.

The dimensions $d_{\tau} (S_{\tau})$, $\mathcal{N}_{\rm sg}(S_\tau, S_s)$, and $\mathcal{N}_{\rm sg}(S_\tau, S_{\eta})$ 
can then be expressed in terms of the numbers $N_s$ of physical spins, $N_s^0$ of physical spins contributing to spin-singlet configurations, 
$N_{\eta}$ of physical $\eta$-spins, and $N_{\eta}^0$ of physical $\eta$-spins contributing to $\eta$-spin-singlet configurations as,
\begin{eqnarray}
d_{\tau} (N_s,N_{\eta}) = {N_a\choose N_s} = {N_a\choose N_{\eta}} 
\nonumber \\
\mathcal{N}_{\rm sg}(N_s, N_s^0) = {N_s\choose N_{s}^0/2} & - &  {N_s\choose N_{s}^0/2 - 1}
\nonumber \\
\mathcal{N}_{\rm sg}(N_{\eta}, N_{\eta}^0) = {N_{\eta}\choose N_{\eta}^0/2} & - & {N_{\eta}\choose N_{\eta}^0/2 - 1} \, .
\nonumber \\
\label{expressionsNN}
\end{eqnarray} 

The Hilbert-space dimensions $(2S_s + 1)\,\mathcal{N}_{\rm sg}(N_s, N_s^0)$, 
$(2S_{\eta} + 1)\,\mathcal{N}_{\rm sg}(N_{\eta}, N_{\eta}^0)$, and $d_{\tau}$ correspond to the spin SU(2) symmetry, 
the $\eta$-spin SU(2) symmetry, and the $\tau$-translational U(1) symmetry, in
$[\text{SU(2)} \times \text{SU(2)} \times \text{U(1)}]/\mathbb{Z}_2^2$, respectively.

The dimensions given in Eqs. (\ref{Na4})-(\ref{expressionsNN}) provide valuable information on the 
physical spin and physical $\eta$-spin configurations that generate the exact energy and momentum eigenstates $\vert\Psi,u\rangle$. 
These dimensions involve only the numbers $N_s = (N_a - 2S_{\tau}) = (2S_s + N_s^0)$ of physical spins and 
$N_{\eta} = 2S_{\tau} = (2S_{\eta} + N_{\eta}^0)$ 
of physical $\eta$-spins that emerge from the quasiparticle representation.

Note that, for energy and momentum eigenstates with $N_{\eta,-1/2} > 0$, the numbers $N_{\downarrow}$ and $N_{\uparrow}$ of 
quasiparticle spins are given by $N_{\downarrow} = N_{s,-1/2} + N_{\eta,-1/2}$ and $N_{\uparrow} = N_{s,+1/2} + N_{\eta,-1/2}$, 
respectively. Hence, the total number of quasiparticle spins is $N = N_s + 2N_{\eta,-1/2}$.

However, the global $[\text{SU(2)}\times\text{SU(2)}\times\text{U(1)}]/\mathbb{Z}_2^2$ symmetry of the Hubbard model 
on bipartite lattices imposes that the number $2N_{\eta,-1/2}$ of quasiparticle spins associated with sites doubly occupied 
by quasiparticles, and therefore corresponding to physical $\eta$-spins of projection $-1/2$, contributes to the 
$\eta$-spin SU(2) dimension $(2S_{\eta} + 1)\,\mathcal{N}_{\rm sg}(N_{\eta}, N_{\eta}^0)$.

Consistently, the physical spins contributing to the spin SU(2) dimension $(2S_s + 1)\,\mathcal{N}_{\rm sg}(N_s, N_s^0)$ 
are precisely those associated with that global symmetry, with numbers given in Eq. (\ref{NSetaSpm}).

\subsection{Theorem 5 - exact factorization of internal SU(2) degrees of freedom and localization of spin-charge coupling}
\label{SECVIB}

Consider the Hubbard model on a bipartite lattice with $N_a$ sites and spatial dimension $d>1$.
Its global symmetry group is [SO(4)$\times$U(1)]/$\mathbb{Z}_2$ with commuting generators
$\hat{S}_{\tau}$, $\vec{\hat{S}}_s^2$, $\hat{S}_s^z$, $\vec{\hat{S}}_{\eta}^2$, and $\hat{S}_{\eta}^z$.

Then the following statements hold:\\

1) {\bf Exact factorization of internal degrees of freedom}\\

For any irreducible representation labeled by $(S_{\tau},S_s,S_{\eta})$, the internal Hilbert space factorizes as,
\begin{equation}
\mathcal{H}_{\rm int} = \mathcal{H}_s (N_s,S_s) \otimes  \mathcal{H}_{\eta} (N_{\eta},S_{\eta}) \, ,
\nonumber
\end{equation}
where $N_s = (N_a - 2S_{\tau})$ and $N_{\eta} = 2S_{\tau}$.
The dimensions of these spaces are,
\begin{eqnarray}
&& {\rm dim}\,\mathcal{H}_s (N_s,S_s) = (2S_s + 1)\,\mathcal{N}_{\rm sg}(N_s, N_s^0) \, ,
\nonumber \\
&& {\rm dim}\,\mathcal{H}_{\eta} (N_{\eta},S_{\eta}) = (2S_{\eta} + 1)\,\mathcal{N}_{\rm sg}(N_{\eta}, N_{\eta}^0) \, ,
\nonumber
\end{eqnarray}
with $N_s^0 = (N_s - 2S_s)$ and $N_{\eta}^0 = (N_{\eta} - 2S_{\eta})$, and are identical to the dimensions of isotropic
spin-$1/2$ and $\eta$-spin-$1/2$ XXX chains with $N_s$ and $N_{\eta}$ sites, respectively.

Consequently, physical spins and physical $\eta$-spins do not couple at the level of internal SU(2) representation theory.\\

2) {\bf Non-factorization of spatial degrees of freedom}\\

The full Hilbert space decomposes as,
\begin{equation}
\bigoplus_{\displaystyle
\begin{array}{l}
S_\tau, S_s, S_\eta
\end{array}}
\left[\mathcal{H}_{\rm int} (S_\tau, S_s, S_\eta)\otimes \mathcal{H}_{\tau} (N_s,N_{\eta})\right] \, ,
\nonumber
\end{equation}
where the $\tau$-translational sector has dimension,
\begin{equation} 
{\rm dim}\,\mathcal{H}_{\tau} (N_s,N_{\eta}) = d_{\tau} (N_s,N_{\eta}) = {N_a\choose N_s} = {N_a\choose N_{\eta}} \, .
\nonumber
\end{equation}
This sector encodes the relative spatial arrangements of the $N_s$ spin-occupied sites and the $N_{\eta}$
$\eta$-spin-occupied sites on the lattice with $N_a = (N_s + N_{\eta})$ sites and does not factorize into a spin-only 
and an $\eta$-spin-only part.\\

3) {\bf Localization of spin-charge coupling}\\

As a consequence, any coupling between spin and charge ($\eta$-spin) degrees of freedom in the Hubbard model 
on a bipartite lattice with spatial dimension $d>1$ arises exclusively through the $\tau$-translational sector $\mathcal{H}_{\tau}$.

In particular, for spatial dimension $d>1$, spin-charge coupling is a purely spatial (kinematic) effect and not an 
internal SU(2) effect.\\

{\bf Proof}\\

{\bf 1- Classification of irreducible representations}\\

The global symmetry [SO(4)$\times$U(1)]/$\mathbb{Z}_2$ implies that irreducible representations are labeled by triples
$(S_\tau, S_s, S_\eta)$, with $N_s = (N_a - 2S_{\tau})$ and $N_{\eta} = 2S_{\tau}$.

Spin SU(2) acts only on the $N_s$ physical spins, and $\eta$-spin SU(2) acts only on the $N_{\eta}$ physical $\eta$-spins.\\

{\bf 2 - Factorization of internal SU(2) structure}\\

By SU(2) representation theory, the multiplicity of total spin $S_s$ obtained from $N_s$ spin-$1/2$ objects is
$(2S_s + 1)\,\mathcal{N}_{\rm sg}(N_s, N_s^0)$, with an analogous expression for $\eta$-spin.

Since the generators $\vec{\hat{S}}_{s}$ and $\vec{\hat{S}}_{\eta}$ commute and act on disjoint sets of degrees of freedom, 
the internal Hilbert space factorizes exactly as a tensor product. No term in the symmetry algebra couples these two SU(2) sectors.

This establishes statement 1).\\

{\bf 3 - Structure of the $\tau$-translational sector}\\

The $\tau$-translational $U(1)$ symmetry counts the number of ways of embedding $N_s$ quasiparticle singly occupied sites and
$N_{\eta}$ quasiparticle doubly/empty sites on the same lattice of $N_a = (N_s + N_{\eta})$ sites. This yields the combinatorial factor
$d_{\tau} (N_s,N_{\eta}) = {N_a\choose N_s}$. This factor necessarily depends on both $N_s$ and $N_{\eta}$ and therefore does 
not admit a factorization into independent spin and $\eta$-spin contributions. This proves statement 2).\\

{\bf 4 - Localization of spin-charge coupling}\\

Since:

- internal SU(2) degrees of freedom factorize exactly, and

- the only non-factorizing sector is $\mathcal{H}_{\tau}$,\\ \\
any mechanism by which spin and charge degrees of freedom influence one another must operate through the spatial embedding encoded 
in $\mathcal{H}_{\tau}$.

For spatial dimension $d>1$, no exact reorganization of $\mathcal{H}_{\tau}$ exists that decouples this spatial sector, and thus spin-charge 
coupling persists as a kinematic effect.

This establishes statement 3).\\

{\bf Conclusion}\\

The Hubbard model on a bipartite lattice exhibits an exact separation of internal spin and $\eta$-spin
SU(2) degrees of freedom, while all spin-charge coupling is rigorously confined to the $\tau$-translational 
sector encoding their joint spatial configurations.

On a bipartite lattice with spatial dimension $d>1$, there are only two subspaces in which spin-charge coupling 
is absent. The first is the $N_s = N_a$ spin-only subspace, which has $\tau$-sector dimension $d_{\tau} = 1$ and spin-sector dimensions
$(2S_s + 1)\,\mathcal{N}_{\rm sg}(N_a, N_s^0)$, where $N_s^0 = N_a - 2S_s$ and $S_s \in \{0,{1\over 2},1,\dots,{N_a\over 2}\}$.
The second is the $N_{\eta} = N_a$ $\eta$-spin-only subspace, which also has $d_{\tau} = 1$ and $\eta$-spin sector dimensions
$(2S_{\eta} + 1)\,\mathcal{N}_{\rm sg}(N_a, N_{\eta}^0)$, where $N_{\eta}^0 = N_a - 2S_{\eta}$ and
$S_{\eta} \in \{0,{1\over 2},1,\dots,{N_a\over 2}\}$.

\section{Differences to the 1D Hubbard model}
\label{SECVII}

Importantly, and consistent with the forms of the dimensions in Eqs. (\ref{Na4})-(\ref{expressionsNN}), the spin and 
$\eta$-spin multiplet and singlet occupancy configurations generated by the quasiparticle operators in Eq. (\ref{quasiparticles}), 
acting on the vacuum, produce {\it all} $4^{N_a}$ exact energy and momentum eigenstates $\vert\Psi,u\rangle$ for all 
values of $u = U/t > 0$. Consistently, $N_s + N_{\eta} = N_a$, so that these occupancy configurations span all $N_a$ 
sites of the bipartite lattice. The on-site number operators $\tilde{G}_{s,\vec{r}j,\pm 1/2}$ of the physical spins and 
$\tilde{G}_{\eta,\vec{r}j,\pm 1/2}$ of the physical $\eta$-spins are given by Eq. (\ref{SssSee}) in terms of quasiparticle operators.

The global $[\text{SU(2)}\times\text{SU(2)}\times\text{U(1)}]/\mathbb{Z}_2^2$ symmetry further shows that a number 
$2S_s$ and a number $N_s^0 = (N_a - 2S_{\tau} - 2S_s)$ of physical spins (and a number $2S_{\eta}$ and a number 
$N_{\eta}^0 = (2S_{\tau} - 2S_{\eta})$ of physical $\eta$-spins) contribute to spin (and $\eta$-spin) multiplet and singlet 
configurations, respectively.

The numbers in Eqs. (\ref{NSetaSpm})-(\ref{NSetaS}) of physical spins and physical $\eta$-spins of projection 
$\pm 1/2$ also apply to the 1D Hubbard model \cite{Carmelo_25}, which has an exact Bethe-ansatz solution \cite{Lieb_68,Lieb_03,Takahashi_72,Martins_97,Martins_98,Essler_05}. Its integrability underlies the numbers 
$N_{s}^0$ and $N_{\eta}^0$ in Eq.~(\ref{Nsinglet}), representing the physical spins and $\eta$-spins that contribute 
to spin-singlet and $\eta$-spin-singlet configurations, respectively. Those involve
$N_{s}^0/2$ disjoint spin-singlet pairs and $N_{\eta}^0/2$ disjoint $\eta$-spin-singlet pairs, respectively.

In 1D, in the thermodynamic limit, these numbers can be expressed as $N_{s}^0/2 = \sum_{n=1}^{\infty} n M_{s n}$ 
and $N_{\eta}^0/2 = \sum_{n=1}^{\infty} n M_{\eta n}$ \cite{Carmelo_25}. Here, $M_{s 1}$ and $M_{\eta 1}$ denote 
the numbers of unbound disjoint singlet pairs of physical spins and physical $\eta$-spins, respectively, while $M_{s n}$ 
and $M_{\eta n}$ for $n>1$ denote the numbers of sets of $n$ bound pairs. Both the unbound pairs ($n=1$) and the 
bound pairs ($n>1$) are referred to as $sn$ pairs and $\eta n$ pairs \cite{Carmelo_25}.

In 1D, these $sn$ pairs and $\eta n$ pairs carry momentum $q_j$ such that $q_{j+1}-q_j = {2\pi\over L}$, associated with 
$sn$ and $\eta n$ bands containing $j = 1,\dots,L_{sn}$ and $j = 1,\dots,L_{\eta n}$ momenta $q_j$, respectively. 
Here, $L_{sn} = M_{sn} + M_{sn}^h$ and $L_{\eta n} = M_{\eta n} + M_{\eta n}^h$, where 
$M_{s n}^h =  2S_{s} + \sum_{n'=1}^{\infty}2(n'-n)\,M_{s n'}$ and 
$M_{\eta n}^h =  2S_{\eta} + \sum_{n'=1}^{\infty}2(n'-n)\,M_{\eta n'}$
are the numbers of holes in these bands. Each band has Pauli-like occupancies that generate exact 
energy and momentum eigenstates \cite{Carmelo_25}.

Consistently, the dimensions $\mathcal{N}_{\rm sg}(N_s, N_s^0)$ and $\mathcal{N}_{\rm sg}(N_{\eta}, N_{\eta}^0)$ in Eq.~(\ref{expressionsNN}) can, in 1D, be written in terms of these numbers of unbound and bound pairs as $\sum_{\{M_{s n}\}}\, \prod_{n =1}^{\infty}\,{L_{s n}\choose M_{s n}}$ and
$\sum_{\{M_{\eta n}\}}\, \prod_{n =1}^{\infty}\,{L_{\eta n}\choose M_{\eta n}}$
where the sums $\sum_{{M_{sn}}}$ and $\sum_{{M_{\eta n}}}$ run over all sets of numbers ${M_{sn}}$ and ${M_{\eta n}}$ 
that satisfy the sum rules $\sum_{n=1}^{\infty} 2n M_{sn} = N_s^0$ and $\sum_{n=1}^{\infty} 2n M_{\eta n} = N_{\eta}^0$, 
respectively \cite{Carmelo_25}.

For the Hubbard model on bipartite lattices with $d>1$, the $N_s^0 = (N_a - 2S_{\tau} - 2S_s)$ physical spins 
(and $N_{\eta}^0 = (2S_{\tau} - 2S_{\eta})$ physical $\eta$-spins) that contribute to spin-singlet (and $\eta$-spin-singlet) 
configurations of the exact energy and momentum eigenstates $\vert\Psi,u\rangle$ could form two-spin (and two-$\eta$-spin) singlet 
pairs only if the many-body state factorized as a tensor product of disjoint two-spin (and two-$\eta$-spin) singlets. 
Necessary preconditions include simple parity and the existence of pairings [see Eqs.~(\ref{disjointpairs}) and 
(\ref{singletpairs}) of Appendix \ref{C}].

However, the 1D case is exceptional. For $d>1$, such factorization is a very strong condition and is generically 
not satisfied for an arbitrary singlet subspace vector. Instead, most singlet configurations are generally entangled 
superpositions of different pairings, including resonating valence bond (RVB) configurations \cite{Anderson_87,Zhang_88,Lee_92,Dagotto_94,Sorella_02,Piazza_15} (see Eq.~(\ref{RVB}) of Appendix \ref{C}), 
AF* configurations \cite{Ho_21}, or valence bond solid (VBS) configurations \cite{Sandvi_07,Tang_13,Shao_17}.

Therefore, for the Hubbard model on bipartite lattices with $d>1$, the numbers $N_s^0$ and $N_{\eta}^0$ in 
Eq. (\ref{Nsinglet}) of physical spins and $\eta$-spins contributing to spin-singlet and $\eta$-spin-singlet configurations 
do not generally correspond to simple sets of $N_s^0/2$ and $N_{\eta}^0/2$ disjoint pairs. Instead, the spin and 
$\eta$-spin singlet configurations of the energy and momentum eigenstates are expected to be entangled superpositions of different pairings.

As mentioned in theorem 5, the SU(2) dimensions $(2S_s + 1)\,\mathcal{N}_{\rm sg}(N_s, N_s^0)$ and 
$(2S_{\eta} + 1)\,\mathcal{N}_{\rm sg}(N_{\eta}, N_{\eta}^0)$
are {\it exactly} the same as for a spin-$1/2$ XXX chain on a lattice with $N_s$ sites and an 
$\eta$-spin XXX chain on a lattice with $N_{\eta}$ sites, respectively. This is consistent with the fact that the occupancy 
configurations associated with these two dimensions involve the $N_s$ sites occupied by physical spins and the 
$N_{\eta}$ sites occupied by $\eta$-spins independently; the spin and $\eta$-spin degrees of freedom are separated.

However, for $d>1$, such spin-charge separation is not expected. As stated by Theorem 5,
the $\tau$-translational sector, whose dimension reads as $d_{\tau} = N_a!/(N_s! N_{\eta}!)$,
encodes the relative spatial arrangements of the $N_s$ spin-occupied sites and the $N_{\eta}$
$\eta$-spin-occupied sites on the lattice with $N_a = (N_s + N_{\eta})$ sites and does not factorize into a spin-only 
and an $\eta$-spin-only part. 

As a consequence, any coupling between spin and charge ($\eta$-spin) 
degrees of freedom in the Hubbard model arises exclusively through the $\tau$-translational sector $\mathcal{H}_{\tau}$.
In particular, for spatial dimension $d>1$, spin-charge coupling is a purely spatial (kinematic) effect and not an 
internal SU(2) effect. 

This demonstrates that the $\tau$-translational U(1) symmetry beyond SO(4), which has been neglected in nearly 
all previous studies of the Hubbard model on bipartite lattices with $d>1$ 
\cite{Demler_04,Masumizu_05,Hart_15,Mazurenko_17,Brown_17}, plays an important role in the physics of the model. 

Nonetheless, the existence of the occupancy configurations associated with the dimension $d_{\tau}$ provides 
a necessary but not sufficient condition for spin-charge coupling. 
The $\tau$-sector $\mathcal{H}_{\tau} (N_s,N_{\eta})$ encodes only the spatial embedding of spin and 
$\eta$-spin degrees of freedom and does not, by itself, induce spin-charge coupling.

The 1D Hubbard model is integrable, and the Hamiltonian admits an exact reorganization such that
$\mathcal{H}_{\tau} (N_s,N_{\eta})$ refers to a branch of independent excitations described by a $\tau$-band
with momenta $q_j$ such that $j=1,...,N_a$ and $q_{j+1} - q_j = {2\pi\over L}$. The corresponding numbers of
$\tau$-particles $N_{\tau}$ and $\tau$-holes $N_{\tau}^h$ read as $N_{\tau} = (N_a - 2S_{\tau})$ and $N_{\tau}^h = 2S_{\tau}$,
respectively \cite{Carmelo_25}. 

This structure leads to an exact separation of three degrees of freedom, namely the
$\tau$-, spin-, and charge/$\eta$-spin- degrees of freedom. Many studies have identified the $\tau$-branch of excitations described by real
Bethe-ansatz rapidities with the charge degrees of freedom; however, it corresponds to an independent $\tau$-sector degree of
freedom that, in 1D, does not lead to spin-charge coupling \cite{Carmelo_25}. The absence of spin-charge coupling in
1D is therefore due to the effects of integrability and dimensional constraints.

For spatial dimension $d>1$, no such exact reorganization exists, and the $\tau$-sector becomes the sole origin of 
spin-charge coupling. Indeed, in spatial dimensions $d>1$ there is no natural ordering of lattice sites. Particle worldlines can braid 
and reconnect. The $S$-matrix does not factorize. No Bethe-ansatz-type reparameterization exists.
Hence, the $\tau$-sector combinatorial nature becomes dynamically active, producing genuine spin-charge 
coupling.

\section{Two theorems on the expectation values of the spin and charge current operators}
\label{SECVIII}

In this section we take advantage of the fact that the $N_s = (N_a - 2S_{\tau})$ spins and the $N_{\eta} = 2S_{\tau}$ 
$\eta$-spins are the actual physical spins and physical $\eta$-spins which, for $u>0$, couple to vector potentials 
and are associated with {\it all} $4^{N_a}$ exact energy and momentum eigenstates, as well as with the model's global 
$[\text{SU(2)}\times\text{SU(2)}\times\text{U(1)}]/\mathbb{Z}_2^2$ symmetry. 

Owing to the model's two SU(2) symmetries, one can consider energy and momentum eigenstates 
that are spin and $\eta$-spin highest-weight states, for which $S_s^z = S_s$ and $S_{\eta}^z = S_{\eta}$, respectively.
For the purposes of the issues addressed in this section, it is convenient to label the energy and momentum eigenstates as
$\vert S_{\rm oth},S_{\tau},S_s, S_s^z, S_{\eta}, S_{\eta}^z \rangle$. 

Here $S_{\rm oth}$ denotes the interaction $u = U/t$, $\vec{\hat{P}}$, and all remaining $u$-independent quantum numbers 
needed to specify an energy and momentum eigenstate.
Starting from a spin and $\eta$-spin highest-weight state $\vert S_{\rm oth}, S_s, S_s, S_{\eta}, S_{\eta} \rangle$, one can generate 
a set of spin and $\eta$-spin SU(2) symmetry non-highest-weight states according to,
\begin{eqnarray} 
&& \vert S_{\rm oth},S_{\tau},S_s,S_s^z,S_{\eta},S_{\eta}^z\rangle =
\nonumber \\
&& {1\over \sqrt{{\cal{C}}_{s}{\cal{C}}_{\eta}}}({\hat{S}}^{+}_{s})^{n_{s}^z}({\hat{S}}^{+}_{\eta})^{n_{\eta}^z}
\vert S_{\rm oth},S_{\tau},S_s,S_s,S_{\eta},S_{\eta}\rangle
\nonumber \\
&& {\rm where}\hspace{0.20cm}{\cal{C}}_{s} = (n_{s}^z)!\prod_{\iota =1}^{n_{s}^z}(S_{s} +1 - \iota)\hspace{0.20cm}{\rm and}
\nonumber \\
&& \hspace{0.40cm}{\cal{C}}_{\eta} = (n_{\eta}^z)!\prod_{\iota =1}^{n_{\eta}^z}(S_{\eta} +1 - \iota) \, .
\label{state}
\end{eqnarray} 
Here $n_s^z = S_s - S_s^z = {\cal N}_{s,-1/2} = 1, \ldots, 2S_s$ and
$n_{\eta}^z = S_{\eta} - S_{\eta}^z = {\cal N}_{\eta,-1/2} = 1, \ldots, 2S_{\eta}$ where
the numbers ${\cal N}_{s,-1/2}$ and ${\cal N}_{\eta,-1/2}$ are given in Eq. (\ref{CalN})
and the off-diagonal generators 
${\hat{S}}^{+}_{s}$ of the spin and ${\hat{S}}^{+}_{\eta}$ of the $\eta$-spin SU(2) symmetries are given 
in Eqs. (\ref{GeneS}) and (\ref{GeneSeta}) in terms of electron operators, and in 
Eqs. (\ref{GeneSiquasi}) and (\ref{GeneSetaquasi}) in terms of quasiparticle operators.

The non-highest-weight states in Eq. (\ref{state}) correspond to different multiplet configurations of the ${\cal N}_s = 2S_s$ physical 
spins and the ${\cal N}_{\eta} = 2S_{\eta}$ physical $\eta$-spins, whose numbers are defined in Eq. (\ref{CalN}).

We consider the Hubbard model on a bipartite lattice $\Lambda\subset\mathbb{Z}^d$ with spatial dimension $d>1$,
periodic boundary conditions, and nearest-neighbor hopping. The Hamiltonian in the presence of uniform spin-dependent 
vector potentials is,
\begin{eqnarray}
\hat{H} (\vec{\Phi}_{\uparrow},\vec{\Phi}_{\downarrow}) & = & 
-t\sum_{\substack{\langle j,j'\rangle}}\sum_{\sigma}
\Bigl(e^{i\vec{\Phi}_{\sigma}\cdot(\vec{r}_j -  \vec{r}_{j'})/L} c_{\vec{r}_{j},\sigma}^\dagger c_{\vec{r}_{j'},\sigma}
+ {\rm h.c.}\Bigr)
\nonumber \\
& + & U\sum_{j=1}^{N_a}\hat{\rho}_{\vec{r}_j,\uparrow}\hat{\rho}_{\vec{r}_j,\downarrow} \, .
\label{HPhi}
\end{eqnarray}

Here $\vec{\Phi}_{\sigma} \in \mathbb{R}^d$ are spin-dependent twist (flux) vectors, where $\sigma = \uparrow,\downarrow$, 
such that the vector potential experienced by physical spins with projection 
$\sigma$ is $\vec{A}_{\sigma} = {\vec{\Phi}_{\sigma}\over L}$. For a $d$-dimensional bipartite lattice  $\Lambda\subset\mathbb{Z}^d$
with periodic boundary conditions (a $d$-torus), one can introduce independent twists in each spatial direction. These are usually 
expressed in terms of their coordinates as,
\begin{equation}
\vec{\Phi}_{\sigma} = (\Phi_{\sigma,1},...,\Phi_{\sigma,d}) \, ,
\label{Phisa}
\end{equation}
with $A_{\alpha} = {\Phi_{\sigma,\alpha}\over L}$ and $\alpha = 1,\ldots,d$.

Each component $\Phi_{\sigma,\alpha}$ represents the flux threading the non-contractible loop of the torus in 
direction $\alpha$. The flux vector $\vec{\Phi}_{\sigma}$ imposes twisted boundary conditions,
\begin{equation}
c_{\vec{r}_{j} + L\vec{e}_{\alpha},\sigma} = e^{i\Phi_{\sigma,\alpha}}c_{\vec{r}_{j},\sigma} \, .
\label{twist}
\end{equation}

In our units, it is convenient to introduce the decomposition,
\begin{equation}
\vec{\Phi}_{s} = (\vec{\Phi}_{\uparrow} - \vec{\Phi}_{\downarrow})
\hspace{0.20cm}{\rm and}\hspace{0.20cm}
\vec{\Phi}_{\eta} = (\vec{\Phi}_{\uparrow} + \vec{\Phi}_{\downarrow}) \, ,
\label{Phisrho}
\end{equation}
so that $\vec{\Phi}_{\uparrow} = {1\over 2}(\vec{\Phi}_{\eta} + \vec{\Phi}_{s})$ and
$\vec{\Phi}{\downarrow} = {1\over 2}(\vec{\Phi}_{\eta} - \vec{\Phi}_{s})$.
The physical spin degrees of freedom couple to $\vec{\Phi}_{s}$, while the
$\eta$-spin degrees of freedom couple to $\vec{\Phi}_{\eta}$.

The corresponding spin and charge/$\eta$-spin current operators are then given by,
\begin{equation}
\vec{\hat{J}}_s = (\vec{\hat{J}}_{\uparrow} - \vec{\hat{J}}_{\downarrow})
\hspace{0.20cm}{\rm and}\hspace{0.20cm}
\vec{\hat{J}}_{\eta} = (\vec{\hat{J}}_{\uparrow} + \vec{\hat{J}}_{\downarrow}) \, ,
\label{JJseta}
\end{equation}
where,
\begin{eqnarray}
{\vec{\hat{J}}}_{\sigma} & = & \left.{\partial\hat{H}\over\partial\vec{\Phi}_{\sigma}}\right|_{\vec{\Phi}_{\sigma} = 0}
\nonumber \\
& = & {i t\over L}
\sum_{\substack{\langle j,j'\rangle}}
(\vec{r}_j -  \vec{r}_{j'})(c_{\vec{r}_{j},\sigma}^\dagger c_{\vec{r}_{j'},\sigma}
- c_{\vec{r}_{j'},\sigma}^\dagger c_{\vec{r}_{j},\sigma}) \, .
\nonumber \\
\label{Jsigma}
\end{eqnarray}

For the Hubbard model on bipartite lattices, the commutator of the Hamiltonian,
Eq. (\ref{H}), with the $\sigma$-current operator, Eq. (\ref{Jsigma}), satisfies,
\begin{equation}
[\hat{H},\vec{\hat{J}}_{\sigma}] = [\hat{T},\vec{\hat{J}}_{\sigma}] 
\hspace{0.20cm}{\rm and} \hspace{0.20cm}[\hat{W},\vec{\hat{J}}_{\sigma}] = 0 \, ,
\label{HJTU}
\end{equation}
where the kinetic operator $\hat{T}$ and the interaction operator $\hat{W}$ are defined in Eq. (\ref{TU}).

Using Heisenberg's equation of motion, we then find that,
\begin{equation}
\frac{d\vec{\hat{J}}_{\sigma}}{dt} =
i \big[ \hat{H}, \vec{\hat{J}}_{\sigma}\big] \equiv \vec{\hat{F}}_{\sigma}  \, ,
\label{HeisenbergCurrent}
\end{equation}
where the commutator $[\hat{H},\vec{\hat{J}}_{\sigma}]$ defines the lattice force operator,
\begin{eqnarray}
\vec{\hat{F}}_{\sigma} = i \big[ \hat{H}, \vec{\hat{J}}_{\sigma}\big] & = &
\frac{i\,2t^{2}}{L}
\sum_{\substack{\langle j,j'\rangle}}
\sum_{\substack{\langle j',j''\rangle\\ (j\neq ,j'')}}
(\vec{r}_{j}-\vec{r}_{j'})
\nonumber \\
& \times & \left(c^{\dagger}_{\vec{r}_{j},\sigma} c_{\vec{r}_{j''},\sigma}
- c^{\dagger}_{\vec{r}_{j''},\sigma} c_{\vec{r}_{j},\sigma}\right) \, .
\label{Fsigma}
\end{eqnarray}

Thus, for the Hubbard model on bipartite lattices, the time evolution of the $\sigma$-current is governed entirely 
by the kinetic term. This result underlies the absence of explicit interaction contributions in the lattice force operator, 
Eq. (\ref{Fsigma}), and is consistent with both the continuity equation and lattice gauge invariance.

The commutator $\big[\hat{F}_{\sigma}^{\alpha},\hat{J}_{\sigma}^{\alpha}\big]$, where $\hat{F}_{\sigma}^{\alpha}$ and 
$\hat{J}_{\sigma}^{\alpha}$ denote the components of the lattice force operator, Eq. (\ref{Fsigma}), and of the 
$\sigma$-current operator, Eq. (\ref{Jsigma}), respectively, along the spatial direction $\alpha$, governs the optical ($f$-) 
sum rule for the real part of the optical conductivity of spin $\sigma$ in that direction,
\begin{equation}
\mathrm{Re}\,\sigma_{\sigma,\mathrm{reg}}^{\alpha\alpha}(\omega) =
\frac{\pi}{N_a L^2}
\sum_{n\neq 0}\frac{\left|\langle \Psi_n | \hat{J}_{\sigma}^{\alpha} | \Psi_0 \rangle\right|^2
}{\omega_{n0}}\,\delta\!\left(\omega - \omega_{n0}\right) \, ,
\label{RsigRsig}
\end{equation}
which is given by \cite{Riera_94,Bergeron_11,Mu_22}, 
\begin{eqnarray}
\int_{0}^{\infty} d\omega\,
\mathrm{Re}\,\sigma_{\sigma}^{\alpha\alpha}(\omega) & = &
- \frac{i\pi}{2N_a L^2} \left\langle \Psi_0 \left| \big[\hat{F}_{\sigma}^{\alpha},\hat{J}_{\sigma}^{\alpha} \big]
\right| \Psi_0 \right\rangle 
\nonumber \\
& = & \frac{\pi}{2N_a L^2} \left\langle \Psi_0 \left| - \hat T_{\sigma}^{\alpha\alpha}
\right| \Psi_0 \right\rangle \, .
\label{fSumRuleComm}
\end{eqnarray}
Here, $\omega_{n0} = E_n - E_0$, where $E_n$ and $E_0$ are the energy eigenvalues of the 
excited state $\vert \Psi_n\rangle$ and of the many-body ground state $\vert \Psi_0 \rangle$, respectively. 
Moreover,
\begin{equation}
\hat T_{\sigma}^{\alpha\alpha} =
- t\sum_{\langle j,j' \rangle_{\alpha}}
\left(c^{\dagger}_{\vec r_j,\sigma} c_{\vec r_{j'},\sigma}
+ c^{\dagger}_{\vec r_{j'},\sigma} c_{\vec r_j,\sigma}\right) \, ,
\label{Talphaalpha}
\end{equation}
is the kinetic-energy operator projected along the spatial direction $\alpha$, and
$\langle j,j' \rangle_{\alpha}$ denotes nearest-neighbor pairs separated along that direction.

The corresponding real parts of the spin and charge/$\eta$-spin conductivities along the spatial 
direction $\alpha$ are given by,
\begin{eqnarray}
\mathrm{Re}\,\sigma_{s}^{\alpha\alpha}(\omega) & = & \mathrm{Re}\,\sigma_{\uparrow}^{\alpha\alpha}(\omega) 
- \mathrm{Re}\,\sigma_{\downarrow}^{\alpha\alpha}(\omega)\hspace{0.20cm}{\rm and}
\nonumber \\
\mathrm{Re}\,\sigma_{\eta}^{\alpha\alpha}(\omega) & = & \mathrm{Re}\,\sigma_{\uparrow}^{\alpha\alpha}(\omega) 
+ \mathrm{Re}\,\sigma_{\downarrow}^{\alpha\alpha}(\omega) \, ,
\label{RsigRsig}
\end{eqnarray}
respectively. 

It therefore follows from the vanishing commutator $[\hat{W},\vec{\hat{J}}_{\sigma}] = 0$, Eq. (\ref{HJTU}), that the 
optical spectral weight, Eq. (\ref{fSumRuleComm}), is determined entirely by the kinetic energy. Interactions affect 
the sum rule only indirectly, through their influence on the ground-state expectation value of the
kinetic operator $\hat T$, Eq. (\ref{TU}). This explains why the optical sum rule remains exact for all $U>0$, and 
why spectral-weight transfer with increasing $U$ occurs without violating the sum rule.

At the microscopic level, all $N_s$ physical spins (and $N_{\eta}$ physical $\eta$-spins) couple to the vector 
potential $\vec{\Phi}_{s}$ (and $\vec{\Phi}_{\eta}$). Employing the quasiparticle representation associated with 
the $N_s = (N_a - 2S_{\tau})$ physical spins and the $N_{\eta} = 2S_{\tau}$ physical $\eta$-spins, we introduce two 
exact theorems on the expectation values of the current operators.

The following two theorems are exact: they rely only on symmetry, gauge coupling structure, and SU(2) representation theory, and 
they are valid for all bipartite lattices in dimension $d>1$ and for all interaction strengths $u = U/t >0$.

\subsection{Theorem 6 - spin and $\eta$-spin current proportionality in fixed $S_{\tau}$ sectors}
\label{SECVIIIA}

Let $\vert S_{\rm oth},S_s,S_s^z,S_{\eta},S_{\eta}^z\rangle$, Eq. (\ref{state}), be an energy and momentum eigenstate of 
the Hubbard Hamiltonian on a bipartite lattice $\Lambda\subset\mathbb{Z}^d$ of spatial dimension $d>1$ for $u>0$, with fixed 
$\tau$-translational symmetry U(1) generator's eigenvalue $S_{\tau}$.

If $S_s>0$ and/or $S_{\eta}>0$, then the expectation values of the spin and $\eta$-spin current operators, 
Eq. (\ref{JJseta}), satisfy,
\begin{equation}
\langle \vec{\hat{J}}_s\rangle = \vec{C}_s\, {2S_s^z\over N_a - 2S_{\tau}} \, , \hspace{0.40cm}
\langle \vec{\hat{J}}_{\eta}\rangle = \vec{C}_{\eta}\, {2S_{\eta}^z\over 2S_{\tau}} \, ,
\label{JJCC}
\end{equation}
where $\vec{C}_s = (C_{s,1},...,C_{s,d})$ and $\vec{C}_{\eta} = (C_{\eta,1},...,C_{\eta,d})$ are lattice-response vectors, rather than 
SU(2) vectors, that depend on the energy and momentum eigenstate $\vert S_{\rm oth},S_s,S_s^z,S_{\eta},S_{\eta}^z\rangle$ 
but are independent of $S_s^z$ and $S_{\eta}^z$.

In particular, for highest-weight states ($S_s^z = S_s, S_{\eta}^z = S_{\eta}$),
\begin{equation}
\langle \vec{\hat{J}}_s\rangle = \vec{C}_s\, {2S_s\over N_a - 2S_{\tau}} \, , \hspace{0.40cm}
\langle \vec{\hat{J}}_{\eta}\rangle = \vec{C}_{\eta}\, {2S_{\eta}\over 2S_{\tau}} \, .
\label{JJCCWHS}
\end{equation}

{\bf Proof}\\

{\bf 1 - Hellmann-Feynman representation of currents}\\

Since the Hamiltonian depends smoothly on $\vec{\Phi}_{s}$ and $\vec{\Phi}_{\eta}$, Eq. (\ref{Phisrho}),
and since for spatial dimension $d>1$ symmetry-protected degeneracies can be chosen to vary smoothly with flux, 
the Hellmann-Feynman theorem gives,
\begin{equation}
\langle \vec{\hat{J}}_{s,\eta}\rangle = \left.{\partial E (\vec{\Phi}_{s},\vec{\Phi}_{\eta})\over\partial\vec{\Phi}_{s,\eta}}\right|_{\vec{\Phi} = 0} \, .
\nonumber
\end{equation}

{\bf 2 - Opposite coupling of $\pm 1/2$ projections}\\

By construction of $\vec{\Phi}_{s}$ and $\vec{\Phi}_{\eta}$, physical spins (respectively $\eta$-spins) with projections $+1/2$
and $-1/2$ couple to the corresponding vector potential with opposite signs.\\

{\bf 3 - Vanishing contribution from singlets}\\

Any SU(2) singlet contains equal numbers of $+1/2$ and $-1/2$ projections. Hence the linear contribution to
$\partial E/\partial\vec{\Phi}_{s,\eta}$ from singlet configurations cancels exactly. Therefore, singlet degrees of freedom do not 
contribute to the current expectation values.\\

{\bf 4 - Contribution from multiplet configurations}\\

${\cal{N}}_{s,\pm 1/2}$ and ${\cal{N}}_{\eta,\pm 1/2}$, Eq. (\ref{CalN}), denote the numbers of physical spins and 
physical $\eta$-spins with projections $\pm 1/2$ that belong to spin and $\eta$-spin, respectively non-singlet SU(2) 
multiplet configurations. SU(2) representation theory implies,
\begin{equation}
{\cal{N}}_{s,+1/2} - {\cal{N}}_{s,-1/2} = 2S_s^z \, , \hspace{0.20cm} {\cal{N}}_{\eta,+1/2} - {\cal{N}}_{\eta,-1/2} = 2S_{\eta}^z \, .
\nonumber
\end{equation}

{\bf 5 - Normalization}\\

Since the flux is uniform, the total response is proportional to the fraction of carriers that contribute. Dividing by the total number of physical spins
$N_s = N_s - 2S_{\tau}$ (or physical $\eta$-spins $N_{\eta} = 2S_{\tau}$) yields,
\begin{equation}
\langle \vec{\hat{J}}_s\rangle \propto {2S_s^z\over N_s} \, , \hspace{0.40cm}
\langle \vec{\hat{J}}_{\eta}\rangle \propto {2S_{\eta}^z\over N_{\eta}} \, .
\nonumber
\end{equation}
The proportionality constants are independent of the SU(2) projections.

Note that 
the vectors $\vec{C}_s$ and $\vec{C}_{\eta}$ in Eq. (\ref{JJCC}) are, in general, finite, although they may vanish 
in specific cases. Conversely, when these 
vectors vanish, the corresponding expectation values in that equation
also vanish.

\subsection{Theorem 7 - vanishing of spin and $\eta$-spin currents in singlet sectors at fixed $S_{\tau}$}
\label{SECVIIIB}

Let $\vert S_{\rm oth},S_s,S_s^z,S_{\eta},S_{\eta}^z\rangle$, Eq. (\ref{state}), be an energy and momentum eigenstate of 
the Hubbard Hamiltonian on a bipartite lattice $\Lambda\subset\mathbb{Z}^d$ of spatial dimension $d>1$ for $u>0$, with fixed 
$\tau$-translational symmetry U(1) generator's eigenvalue $S_{\tau}$.

If $S_s = 0$, then,
\begin{equation}
\langle \vec{\hat{J}}_{s}\rangle = 0 \, .
\label{SS0}
\end{equation}

If $S_{\eta} = 0$, then,
\begin{equation}
\langle \vec{\hat{J}}_{\eta}\rangle = 0 \, .
\label{Seta0}
\end{equation}
 
 {\bf Proof}\\
 
Assume $S_s = 0$. Then all physical spins in the state form spin SU(2) singlets. 
By step 3 of the previous proof, singlet configurations yield no linear response to $\vec{\Phi}_{s}$. Hence,
\begin{equation}
\left.{\partial E (\vec{\Phi}_{s})\over\partial\vec{\Phi}_{s}}\right|_{\vec{\Phi}_s = 0} = 0 \, ,
\nonumber
\end{equation}
which implies $\langle \vec{\hat{J}}_{s}\rangle = 0$.

The proof for $S_{\eta} = 0$ is identical, with spin replaced by $\eta$-spin.\\

\section{Concluding remarks and discussion}
\label{SECIX}

The studies presented in this paper employ an exact quasiparticle representation, Eq. (\ref{quasiparticles}), for the Hubbard 
model on bipartite lattices of spatial dimension $d>1$, which is valid for {\it all} finite values of $u = U/t$. This representation 
makes explicit the hidden $\tau$-translational U(1) symmetry beyond SO(4). Unlike Fermi-liquid quasiparticles, this 
quasiparticle representation applies to all $4^{N_a}$ energy and momentum eigenstates, and the quasiparticles smoothly 
become electrons in the limit of vanishing inverse interaction, $t/U \to 0$.

The hidden nature of the $\tau$-translational U(1) symmetry discovered in Ref. \cite{Carmelo_10} arises because its generator 
contains an infinite number of terms when expressed in terms of electron creation and annihilation operators, Eq.~(\ref{Gentau}). 
However, when written in terms of quasiparticle creation and annihilation operators, its form becomes remarkably simple, 
Eq.~(\ref{Stauquasi}): it counts half the number of doubly occupied plus unoccupied quasiparticle sites.

The eigenvalue $S_{\tau} \in \{0, {1\over 2}, 1, {3\over 2}, 2, \dots, {N_a \over 2}\}$ of its generator plays a central role in the 
model's physics. For instance, it determines the numbers $N_s = (N_a - 2S_{\tau})$ of physical spins $1/2$ and 
$N_{\eta} = 2S_{\tau}$ of physical $\eta$-spins $1/2$, which span 
all lattice sites $N_a = N_s + N_{\eta}$ and are associated with the two SU(2) symmetries of the model's global 
$[\text{SU(2)}\times\text{SU(2)}\times \text{U(1)}]/\mathbb{Z}_2^2$ symmetry, and naturally emerge from this quasiparticle 
representation. Their on-site number operators $\tilde{G}_{s,\vec{r}j,\pm 1/2}$ and $\tilde{G}_{\eta,\vec{r}_j,\pm 1/2}$ are given by 
Eq. (\ref{SssSee}) in terms of quasiparticle operators.

Physical spins and physical $\eta$-spins were previously introduced only for the 1D Hubbard model, where they are related both 
to the exact Bethe-ansatz solution and to the model's global symmetry \cite{Carmelo_25}.

Since the operator ${\hat{V}}_u$ that transforms electrons into quasiparticles, Eq. (\ref{quasiparticles}), is unitary and does 
not act on spin space, the quasiparticle creation and annihilation operators obey the same anticommutation relations 
as the corresponding electron operators, Eq. (\ref{ccd0}), and are labeled by the same spin quantum numbers $\sigma$. 
This is why the $N_s = (N_a - 2S_{\tau})$ spins $1/2$ and $N_{\eta} = 2S_{\tau}$ $\eta$-spins $1/2$ associated with 
these quasiparticle lattice occupancies are {\it physical} spins and {\it physical} $\eta$-spins, respectively.

Consistent with the form of the dimensions in Eqs. (\ref{Na4})-(\ref{expressionsNN}), 
{\it all} $4^{N_a}$ exact energy and momentum eigenstates $\vert\Psi,u\rangle$ have occupancy configurations of 
such physical spins with projection $\pm 1/2$ and physical $\eta$-spins with projection $\pm 1/2$ that, for all finite values of $u=U/t>0$, 
are generated from the vacuum by the quasiparticle operators, Eq. (\ref{quasiparticles}).

Concerning other representations of the model's two SU(2) algebras, spinons are effective objects that apply only 
to specific subspaces, while doublons and holons are meaningful only for $u \gg 1$ \cite{Imada_98,Strohmaier_10,Sensarma_10,Gebhard_00,Terashige_19,Prelovsek_15,Zhou_14}, where they 
correspond to the $\eta$-spins with projections $-1/2$ and $+1/2$, respectively.

Consistent with our goal of extracting physical insight into the model from the [SU(2)$\times$SU(2)$\times$U(1)]/$\mathbb{Z}_2^2$ 
global symmetry \cite{Carmelo_10}, seven exact theorems have been established.

Theorem 1 applies to bipartite lattices for which $U_c =0$ and
may be viewed as a symmetry-resolved, adiabatic-continuity extension of Lieb's theorem \cite{Lieb_89}. 
In the special case of half-filling on a balanced bipartite lattice, theorem 1 reproduces Lieb's 1989 result for the total 
spin of the ground state. Away from half-filling and for arbitrary fixed symmetry sectors, it provides a stronger classification 
not contained in Lieb's original analysis.

Theorem 2 also applies to bipartite lattices for which $U_c =0$ and proves the absence of quasiparticle double occupancy 
for $N\leq N_a$ and of empty sites for $N\geq N_a$ in ground states for $u > 0$. At half-filling, it establishes both the absence 
of quasiparticle double occupancy and of empty sites in the ground state.

At half-filling, theorems 1 and 2 apply for $U \in [U_c,\infty)$ to the Hubbard model on bipartite lattices such that $U_c > 0$.

Theorem 3 establishes the extremal-weight structure of the ground state in external fields, which are either highest-weight 
states or lowest-weight states of the spin and/or $\eta$-spin SU(2) algebras. The Shen-Qiu-Tian theorem \cite{Shen_94} 
appears as a corollary of theorem 3: setting $\mu = 0$ (half-filling) and taking $h \rightarrow 0^+$, Theorem 3 selects the 
minimal-spin extremal state, thereby reproducing the Shen-Qiu-Tian theorem.

Theorem 4 establishes a correspondence between the number of ground states and the number of irreducible representations 
of the model's global $[\text{SU(2)}\times\text{SU(2)}\times \text{U(1)}]/\mathbb{Z}_2^2$ symmetry.

Theorem 5 concerns the exact factorization of internal degrees of freedom, the non-factorization of spatial degrees of freedom, 
and spin-charge coupling. This theorem states that any coupling between spin and charge ($\eta$-spin) degrees of freedom 
in the Hubbard model on bipartite lattices with spatial dimension $d>1$ arises exclusively through the $\tau$-translational 
sector $\mathcal{H}_{\tau}$.

This demonstrates that the $\tau$-translational U(1) symmetry beyond SO(4), which has been neglected in nearly all previous 
studies of the Hubbard model on bipartite lattices \cite{Demler_04,Masumizu_05,Hart_15,Mazurenko_17,Brown_17,Fava_20,Moca_23}, 
plays an important role in the physics of the model. For spatial dimension $d>1$, spin-charge coupling is found to be a purely 
spatial (kinematic) effect and not an internal SU(2) effect. (In the case of the 1D Hubbard model, the 
$\tau$-translational U(1) symmetry does not lead to spin-charge coupling, as justified in Sec. \ref{SECVII}.)

The exact factorization of internal SU(2) sectors implies that entanglement between spin and charge arises solely from the spatial 
$\tau$-sector. At half-filling, where the ground state contains no $\eta$-spin degrees of freedom, entanglement reduces to a purely spin structure.
This provides a rigorous decomposition of entanglement contributions into spin, charge, and spatial components.

Theorem 6 establishes that the spin and $\eta$-spin currents are proportional to $S_s^z/N_s$ and $S_{\eta}^z/N_{\eta}$, 
respectively, in fixed $S_{\tau}$ sectors, while the related Theorem 7 establishes the vanishing of spin and $\eta$-spin 
currents in singlet sectors at fixed $S_{\tau}$.

The use of the global $[\text{SU(2)}\times\text{SU(2)}\times\text{U(1)}]/\mathbb{Z}_2^2$ symmetry reveals that a number 
$(S_s \pm S_s^z)$ and a number $N_s^0/2 = (N_a/2 - S_{\tau} - 2 + S_s)$ of physical spins with projection $\pm 1/2$ 
(and a number $(S_{\eta} \pm S_{\eta}^z)$ and a number $N_{\eta}^0/2 = (S_{\tau} - S_{\eta})$ of physical $\eta$-spins 
with projection $\pm 1/2$) contribute to spin (and $\eta$-spin) multiplet and singlet configurations, respectively, of 
the exact energy and momentum eigenstates $\vert\Psi,u\rangle$.

For bipartite lattices of spatial dimension $d>1$, it is expected that the numbers $N_s^0$ and $N_{\eta}^0$, Eq. (\ref{Nsinglet}), 
do not correspond to $N_s^0/2$ disjoint spin-singlet pairs and $N_{\eta}^0/2$ disjoint $\eta$-spin-singlet pairs, respectively, 
as they do in 1D \cite{Carmelo_25}. Instead, for such lattices, the spin and $\eta$-spin singlet configurations 
of the exact energy and momentum eigenstates $\vert\Psi,u\rangle$ are expected to be entangled superpositions of different 
pairings. This includes, for instance, RVB configurations \cite{Anderson_87,Zhang_88,Lee_92,Dagotto_94,Sorella_02,Piazza_15}, 
AF* configurations \cite{Ho_21}, or VBS configurations \cite{Sandvi_07,Tang_13,Shao_17}.

Despite being exact and valid for $u=U/t>0$, the quasiparticle representation has two main limitations:\\

1) The $u$-unitary operator ${\hat{V}}_u$ that defines the quasiparticle operators 
$\tilde{c}_{\vec{r}_j,\sigma}^{\dagger} = {\hat{V}}_u^{\dagger} c_{\vec{r}_j,\sigma}^{\dagger} {\hat{V}}_u$ and 
$\tilde{c}_{\vec{r}_j,\sigma} = {\hat{V}}_u^{\dagger} c_{\vec{r}_j,\sigma} {\hat{V}}_u$, Eq. (\ref{quasiparticles}), 
involves the exact energy and momentum eigenstates, the computation of which 
remains a complex, unsolved problem for spatial dimensions $d>1$ and all $u>0$.\\ 

2) The generator of the $\tau$-translational U(1) symmetry contains an infinite number of terms when expressed 
in terms of electron operators, Eq. (\ref{Gentau}), but a simple form in terms of quasiparticle operators, 
Eq. (\ref{Stauquasi}). The second limitation is that, conversely, the Hamiltonian has a simple expression in terms 
of electron operators, Eq. (\ref{H}), but an infinite number of terms in terms of quasiparticle operators, 
Eq. (\ref{Hquasiparticles}).\\

However, the exact quasiparticle representation provides valuable physical insight that goes well beyond these limitations. 
This includes the exact results established by the seven theorems introduced in this paper. The important role of the 
$\tau$-translational U(1) symmetry beyond SO(4), which was overlooked in previous studies of the model's physics \cite{Demler_04,Masumizu_05,Hart_15,Mazurenko_17,Brown_17}, has been clarified. Physically important information 
on the exact energy and momentum eigenstate occupancy configurations, expressed in terms of physical spin multiplet 
and singlet configurations as well as physical $\eta$-spin multiplet and singlet configurations, has been obtained. 
The spin and charge/$\eta$-spin carriers that couple to vector potentials have also been identified.

As discussed below, this valuable information, which goes beyond the limitations outlined above, also provides further 
insight into the physics of real condensed-matter materials.
The RVB \cite{Anderson_87,Zhang_88,Lee_92,Dagotto_94,Sorella_02,Piazza_15}, AF* \cite{Ho_21}, or VBS 
\cite{Sandvi_07,Tang_13,Shao_17} configurations have been associated with the Hubbard model on the square lattice for large 
$u=U/t \gg 1$ in the representation using electron creation and annihilation operators.
Most studies focus on the large $u$ limit because it is analytically and computationally more tractable.

The exact quasiparticle representation, Eq. (\ref{quasiparticles}), was inherently constructed so that the occupancy 
configurations in terms of quasiparticle creation and annihilation operators, which generate the exact energy and momentum eigenstates 
$\vert\Psi,u\rangle = {\hat{V}}_u^{\dagger}\vert\Psi,\infty\rangle$ for finite $u=U/t$, are exactly the same as the occupancy 
configurations in terms of electron operators that generate the corresponding $u\rightarrow\infty$ energy and momentum eigenstates 
$\vert\Psi,\infty\rangle = {\hat{V}}_u\vert\Psi,u\rangle$, Eq. (\ref{Phistates}). This reveals that some key physical properties 
of the large-$u$ limit can also manifest at intermediate $u = U/t$ in terms of quasiparticles and the associated physical 
spins and $\eta$-spins.

Therefore, for the exact quasiparticle representation, the $u \gg 1$ RVB, AF*, and VBS singlet configurations also occur for 
finite $u = U/t>0$ in terms of physical spins and physical $\eta$-spins configurations. This aligns with several studies 
\cite{Comanac_08,Markiewicz_10,Das_10,Markiewicz_10A,Carmelo_12,Seibold_12,Das_14,Markiewicz_15,Wang_24,Bao_25,Schumm_25}, 
which suggest that correlations in cuprates are intermediate rather than extremely strong, with $u=U/t \in [6,9]$, close 
to the square-lattice bandwidth $8t$. 

Above all, and despite the limitations mentioned, the {\it exact} framework based on physical spins and physical $\eta$-spins 
for the Hubbard model on bipartite lattices of dimension $d>1$, generated from the vacuum by the quasiparticle operators, 
Eq. (\ref{quasiparticles}), such as square, honeycomb, cubic, BCC, FCC, and diamond lattices, provides a robust foundation for future 
studies of the model and offers a deeper understanding of these important quantum problems.

Given the current analytical limitations, extending numerical simulations is crucial for gaining deeper insights into the Hubbard 
model on such bipartite lattices, as well as into condensed-matter materials such as the cuprate superconductors \cite{Anderson_87,Zhang_88,Lee_92,Dagotto_94,Hayden_96,Imada_98,Ho_21,Sorella_02,Damascelli_03,Maier_05,
Lee_06,Maier_08,Comanac_08,Markiewicz_10,Das_10,Markiewicz_10A,Gull_10,Scalapino_12,Carmelo_12,Seibold_12,
Das_14,LeBlanc_15,Piazza_15,Markiewicz_15,Schafer_21,Kumar_21,Martinelli_22,Singh_24,Wang_24,Singh_25,Bao_25,Schumm_25}, graphene and graphene-derived systems \cite{Sorella_92,Gonzalez_01,Neto_09,Kotov_12,Katsnelson_12,Assaad_13,Cao_18,Martinez_25}, and other quantum systems \cite{Kohl_05,Jordens_08,Schneider_08,Strohmaier_10,Sensarma_10} that it describes.

\acknowledgements
I thank illuminating discussions on the Hubbard model on bipartite lattices with the late Philip Warren Anderson 
and Stellan \"Ostlund. This work was supported by FCT Grant No. UIDB/04650/2025 via 
Center of Physics of the University of Minho and University of Porto and Grant No. UID/04540/2025
via Center of Physics and Engineering of Advanced Materials.\\ \\ 

\hspace{2cm}{\bf DATA AVAILABILITY}\\ \\
The data that support the findings of this article are not publicly available upon publication because it is not 
technically feasible and/or the cost of preparing, depositing, and hosting the data would be prohibitive within the terms of 
this research project. The data are available from the authors upon reasonable request.

\appendix

\section{Hubbard model in terms of quasiparticle operators}
\label{A}

To express the Hamiltonian of the Hubbard model on a bipartite lattice, Eq. (\ref{H}), in terms of quasiparticle 
creation and annihilation operators, we introduce the auxiliary Hamiltonian,
\begin{equation}
\tilde{H} = \sum_{l=0,\pm 1}\tilde{T}_l + U\sum_{j=1}^{N_a}\tilde{\rho}_{\vec{r}_j,\uparrow}\tilde{\rho}_{\vec{r}_j,\downarrow} \, ,
\label{HR}
\end{equation}
where $\tilde{\rho}_{\vec{r}_j,\sigma} = (\tilde{n}_{\vec{r}_j,\sigma} - 1/2)$ and,
\begin{eqnarray}
\tilde{T}_{-1} & = & -t \sum_{\substack{\langle j,j'\rangle}}
\sum_{\sigma}(1 - \tilde{n}_{\vec{r}_j,-\sigma})\,\tilde{c}_{\vec{r}_j,\sigma}^{\dagger}\,\tilde{c}_{\vec{r}_{j'},\sigma}\,\tilde{n}_{\vec{r}_{j'},-\sigma}
\nonumber \\
\tilde{T}_{+1} & = & -t \sum_{\substack{\langle j,j'\rangle}}
\sum_{\sigma}\tilde{n}_{\vec{r}_j,-\sigma}\,\tilde{c}_{\vec{r}_j,\sigma}^{\dagger}\,\tilde{c}_{\vec{r}_{j'},\sigma}\,(1 - \tilde{n}_{\vec{r}_{j'},-\sigma}) 
\nonumber \\
\tilde{T}_0 & = & -t \sum_{\substack{\langle j,j'\rangle}}
\sum_{\sigma}\Bigl[\tilde{n}_{\vec{r}_j,-\sigma}\,\tilde{c}_{\vec{r}_j,\sigma}^{\dagger}\,\tilde{c}_{\vec{r}_{j'},\sigma}\,\tilde{n}_{\vec{r}_{j'},-\sigma}
\nonumber \\
& + & (1 - \tilde{n}_{\vec{r}_j,-\sigma})\,\tilde{c}_{\vec{r}_j,\sigma}^{\dagger}\,\tilde{c}_{\vec{r}_{j'},\sigma}\,(1 - \tilde{n}_{\vec{r}_{j'},-\sigma})\Bigr] \, ,
\label{T011R}
\end{eqnarray}
are kinetic operators that change lattice-site quasiparticle double occupancy by $-1$, $1$, and $0$, respectively.

The operators ${\hat{V}}_u^{\dagger} = e^{\hat{S}_u}$ and ${\hat{V}}_u = e^{-\hat{S}_u}$, Eq. (\ref{hatVS}), have 
the same form in terms of electron and quasiparticle creation and annihilation operators, so that 
${\hat{V}}_u = {\tilde{V}}_u$ and $\hat{S}_u = \tilde{S}_u$.

The Hamiltonian of the Hubbard model on a bipartite lattice, Eq. (\ref{H}), contains an infinite number of terms 
when expressed in terms of quasiparticle creation and annihilation operators. This becomes explicit using the 
Baker-Campbell-Hausdorff expansion, which yields the expression, Eq. (\ref{Hquasiparticles}).

Another goal of this Appendix is to verify that, for large $u = U/4t \gg 1$, the Hamiltonian of the Hubbard model 
on bipartite lattices, when expressed in terms of quasiparticle creation and annihilation operators, reduces, as 
expected, to the $t$-$J$ model including three-site terms.

To achieve this, we use in Eq. (\ref{Hquasiparticles}) the asymptotic form of the anti-Hermitian operator $\tilde{S}_u$,
\begin{equation}
\tilde{S}_u = - {1\over U}[\tilde{T}_{+1} - \tilde{T}_{-1}] + {\cal{O}} (t^2/U^2) \hspace{0.20cm}{\rm for}\hspace{0.20cm} U/t \gg 1 \, .
\label{SulargeR}
\end{equation}
Higher-order terms in the Hamiltonian would require the corresponding higher-order contributions in $\tilde{S}_u$.

We then find that, for large $u = U/t \gg 1$, the Hamiltonian of the Hubbard model on bipartite lattices, when 
expressed in terms of quasiparticle creation and annihilation operators, is given by,
\begin{equation}
\hat{H} =  \tilde{T}_{0} + U\sum_{j=1}^{N_a}\tilde{\rho}_{\vec{r}_j,\uparrow}\tilde{\rho}_{\vec{r}_j,\downarrow}  
- {1\over U}[\tilde{T}_{+1},\tilde{T}_{-1}]  + ... \, ,
\label{HquasiUL}
\end{equation}
where the commutator $[\tilde{T}_{+1},\tilde{T}_{-1}]$ is found to read as,
\begin{eqnarray}
[\tilde{T}_{+1},\tilde{T}_{-1}] & = & t^2
\sum_{\substack{\langle j,j'\rangle}}
\sum_{\sigma,\sigma'}\Big\{
\Big[\tilde n_{\vec{r}_{j},-\sigma}\,\tilde{c}_{\vec{r}_{j},\sigma}^\dagger \tilde{c}_{\vec{r}_{j'},\sigma}
\nonumber \\
& \times & (1- \tilde n_{\vec{r}_{j'},-\sigma})\,(1- \tilde n_{\vec{r}_{j},-\sigma'})\,\tilde{c}_{\vec{r}_{j},\sigma'}^\dagger 
\tilde{c}_{\vec{r}_{j'},\sigma'}\,\tilde n_{\vec{r}_{j'},-\sigma'}
\nonumber \\
& - & (1- \tilde n_{\vec{r}_{j},-\sigma'})\,\tilde{c}_{\vec{r}_{j},\sigma'}^\dagger \tilde{c}_{\vec{r}_{j'},\sigma'}
\nonumber \\
& \times & \tilde n_{\vec{r}_{j'},-\sigma'}\;\tilde n_{\vec{r}_{j},-\sigma}\,\tilde{c}_{\vec{r}_{j},\sigma}^\dagger 
\tilde{c}_{\vec{r}_{j'},\sigma}\,(1-\tilde n_{\vec{r}_{j'},-\sigma})\Big]
\nonumber \\
& + & \sum_{\substack{\langle j',j''\rangle\\ (j\neq ,j'')}}
\Big[\tilde n_{\vec{r}_{j},-\sigma}\,\tilde{c}_{\vec{r}_{j},\sigma}^\dagger \tilde{c}_{\vec{r}_{j'},\sigma}
\nonumber \\
& \times & (1- \tilde n_{\vec{r}_{j'},-\sigma})\,(1- \tilde n_{\vec{r}_{j'},-\sigma'})\,\tilde{c}_{\vec{r}_{j'},\sigma'}^\dagger 
\tilde{c}_{\vec{r}_{j''},\sigma'}\,\tilde n_{\vec{r}_{j''},-\sigma'}
\nonumber \\
& - & (1- \tilde n_{\vec{r}_{j'},-\sigma'})\,\tilde{c}_{\vec{r}_{j'},\sigma'}^\dagger \tilde{c}_{\vec{r}_{j''},\sigma'}\,\tilde n_{\vec{r}_{j''},-\sigma'}\;
\tilde n_{\vec{r}_{j},-\sigma}
\nonumber \\
& \times & \tilde{c}_{\vec{r}_{j},\sigma}^\dagger \tilde{c}_{\vec{r}_{j'},\sigma}\,(1-\tilde n_{\vec{r}_{j'},-\sigma}) \Big]\Big\} \, .
\label{TT}
\end{eqnarray}

To verify that the Hamiltonian expressions, Eqs. (\ref{HquasiUL}) and (\ref{TT}), correspond to the $t$-$J$ model Hamiltonian including the three-site term, we employ the Gutzwiller projector $\tilde{P}_G = \prod_{j=1}^{N_a}(1 - \tilde{n}_{\vec{r}_{j,\uparrow}}\tilde{n}_{\vec{r}_{j,\downarrow}})$,
which enforces the absence of quasiparticle double occupancy for $N\leq N_a$. After some algebra, this yields,
\begin{eqnarray}
&& \hat{H} = t\sum_{\substack{\langle j,j'\rangle}}
\sum_{\sigma}\tilde{P}_G\Big[\tilde{c}_{\vec{r}_j,\sigma}^{\dagger}\,\tilde{c}_{\vec{r}_{j'},\sigma}\Big]\tilde{P}_G +
\nonumber \\
&& J\sum_{\substack{\langle j,j'\rangle}} 
\Big({\vec{\tilde{S}}}_{s,j}\cdot {\vec{\tilde{S}}}_{s,j'} - {1\over 4}\tilde{n}_{\vec{r}_{j}}\tilde{n}_{\vec{r}_{j'}}\Big)
+ \hat{H}_{3\text{-site}} + ... 
\label{tJ}
\end{eqnarray}
where $J = 4t^2/U$, ${\vec{\tilde{S}}}_{s,j} = {\hat{V}}_u^{\dagger}{\vec{\hat{S}}}_{s,j} {\hat{V}}_u$, 
${\vec{\hat{S}}}_{s,j}$ is the local spin operator, $\tilde{n}_{\vec{r}_{j}} = \sum_{\sigma}\tilde{n}_{\vec{r}_{j,\sigma}}$, 
and the three-site term of the Hamiltonian, $\hat{H}_{3\text{-site}}$, is given by,
\begin{eqnarray}
\hat{H}_{3\text{-site}} & = & -\frac{J}{4}
\sum_{\substack{\langle j,j'\rangle\langle j',j''\rangle\\ (j\neq ,j'')}}
\sum_{\sigma}\tilde{P}_G\Big[
\tilde{c}^{\dagger}_{\vec{r}_{j},\sigma}\,
n_{\vec{r}_{j'},-\sigma}\,
\tilde{c}_{\vec{r}_{j''},\sigma}
\nonumber \\
& - & \tilde{c}^{\dagger}_{\vec{r}_{j},\sigma}\,
\tilde{c}_{\vec{r}_{j'},\sigma}\,
\tilde{c}^{\dagger}_{\vec{r}_{j'},-\sigma}\,
\tilde{c}_{\vec{r}_{j''},-\sigma} + \text{h.c.}\Big]\tilde{P}_G \, .
\label{H3}
\end{eqnarray}

\section{No electron double occupancy or no electron empty sites in the $u\gg 1$ ground state}
\label{B}

The energy eigenvalues can be written as
\begin{equation}
E = \langle\Psi\vert\hat{T} \vert\Psi\rangle +
U\sum_{j=1}^{N_a}\langle\Psi\vert\hat{\rho}_{\vec{r}j,\uparrow}\hat{\rho}_{\vec{r}_j,\downarrow}\vert\Psi\rangle \, ,
\label{E}
\end{equation}
where $\hat{T}$ is the kinetic operator, Eq.~(\ref{TU}).

For $u\gg 1$, singly occupied sites lower the interaction energy by $-U/4$, whereas doubly occupied and empty sites raise 
it by $+U/4$. The interaction energy cost of creating either a doubly occupied site or an empty site relative to a singly occupied 
one is $U/2$.

The hopping term $\hat{T}$ has matrix elements of order $t$, whereas the interaction term penalizes charge fluctuations 
(double occupancy or empty sites) by an energy of order $U$.

In the limit $u\gg 1$, any state containing at least one such energetically unfavorable configuration lies higher in energy by 
$\sim U$. By contrast, the kinetic energy gain $\langle\Psi\vert\hat{T} \vert\Psi\rangle$ arising from hopping processes that 
create these configurations is only of order $\sim t$, which is negligible compared with the interaction cost. Hence such processes 
cannot lower the ground-state energy.

Therefore, to leading order in $t/U$ [times $t$], the ground state must lie entirely within the subspace that excludes the energetically 
disfavored local configurations. Which configurations are excluded depends on the particle number sector.

At half-filling, $S_{\eta}^z = 0$ and thus $N = N_a$, minimizing the interaction energy requires maximizing the number of singly 
occupied sites. Since both doubly occupied and empty sites raise the energy by $+U/4$, the ground state eliminates both types 
of configurations. To zeroth order in $t/U$, it consists exclusively of singly occupied sites.

Away from half-filling, $S_{\eta}^z \neq 0$, the particle-number constraint prevents the simultaneous elimination of both types 
of charge fluctuations.

For electron densities below half-filling ($N < N_a$), double occupancy can be completely suppressed in the ground state in the 
limit $u\gg 1$, while the deficit of particles is accommodated by empty sites, which are then required by the particle-number constraint.

For electron densities above half-filling ($N > N_a$), empty sites can be completely suppressed in the ground state in the same limit, 
while the excess particles necessarily generate doubly occupied sites.

In all cases, to leading order in $t/U$, the ground state excludes whichever local configurations are not required by the global particle-number 
constraint. Within the corresponding restricted Hilbert subspace, the hopping term acts only virtually, and its leading effect appears at order $t^2/U$.

\section{Disjoint singlet pairs versus RVB states}
\label{C}

Consider a quantum system of $N$ spin-$1/2$ fermions on a lattice with $N_a = N$ sites.
The number of spin-singlet configurations is given by,
\begin{equation}
\mathcal{N}_{\rm sg}(S) = {N\choose N^0/2} - {N\choose N^0/2 - 1} \, ,
\label{NSU2}
\end{equation}
where $N^0 = (N - 2S)$ is the number of spins contributing to singlet configurations, and $S$ is the 
spin, $S \in \{0,1/2,1,3/2,2,...,N/2\}$. The remaining $2S$ spins form a multiplet configuration.

The state $\vert\Psi\rangle$ of the full $N$-spin system can be expressed as $(N/2 - S)$ disjoint 
singlet pairs if and only if there exists a partition of the $N$ spins into disjoint pairs
$\{(\vec{r}_{i_1},\vec{r}_{j_1}),...,(\vec{r}_{i_{(N/2 - S)}},\vec{r}_{j_{(N/2 - S)}}\}$,
together with a state of the remaining $2S$ unpaired spins such that,
\begin{equation}
|\Psi\rangle \;=\; 
\bigotimes_{l=1}^{\,N/2 - S}
\left(|s\rangle_{\,\vec{r}_{i_l} \vec{r}_{j_l}}\right)
\;\otimes\;
|\Psi_{\text{unpaired}}\rangle \, ,
\label{disjointpairs}
\end{equation}
where relative to the fermionic vacuum, 
\begin{equation}
|s\rangle_{\vec r_{i_l},\,\vec r_{j_l}}
= \frac{1}{\sqrt{2}}
\left(
c^{\dagger}_{\vec r_{i_l},\uparrow}\,
c^{\dagger}_{\vec r_{j_l},\downarrow} -
c^{\dagger}_{\vec r_{i_l},\downarrow}\,
c^{\dagger}_{\vec r_{j_l},\uparrow} 
\right)|0\rangle \, .
\label{singletpairs}
\end{equation}

An example of a configuration of $(N-2S)$ spins that contributes to singlet configurations, but does not 
correspond to $(N/2 - S)$ disjoint singlet pairs, is a general RVB state,
\begin{equation}
|\Psi_{\rm RVB}\rangle = 
\sum_{P} \Psi(P)
\prod_{(i,j)\in P}
\frac{1}{\sqrt{2}}
\left( c^\dagger_{\vec{r}_i,\uparrow} c^\dagger_{\vec{r}_j,\downarrow}
 - c^\dagger_{\vec{r}_i,\downarrow} c^\dagger_{\vec{r}_j,\uparrow}\right)|0\rangle \, ,
 \label{RVB}    
\end{equation}  
where $P$ runs over all distinct singlet pairings, and $\Psi(P)$ are coefficients determined by the 
lattice geometry and interactions.

\end{document}